\documentclass[preprint,12pt,authoryear]{elsarticle}
\usepackage[outline]{contour}
\usepackage{graphicx,amssymb,amsfonts,amsmath,color,rotating,multirow,array,setspace}
\usepackage[percent]{overpic}
\makeatletter
\def\ps@pprintTitle{%
 \let\@oddhead\@empty
 \let\@evenhead\@empty
 \def\@oddfoot{\centerline{\thepage}}%
 \let\@evenfoot\@oddfoot}
\makeatother

\makeatletter
\g@addto@macro{\endtabular}{\rowfont{}}
\makeatother
\newcommand{\rowfonttype}{}
\newcommand{\rowfont}[1]{
 \gdef\rowfonttype{#1}#1%
}
\newcolumntype{L}{>{\rowfonttype}l}

\begin{document}

\begin{frontmatter}



\title{A Look at Financial Dependencies by Means of Econophysics and Financial Economics}


\author[label1,label3]{M. Raddant}
\ead{raddant@csh.ac.at}
\author[label4,label3,label5]{T. Di Matteo}
\ead{tiziana.di\_matteo@kcl.ac.uk}
\address[label1]{University for Continuing Education Krems, Department for Knowledge and\\ Communication Management, 3500 Krems, Austria}
\address[label3]{Complexity Science Hub Vienna, Josefst{\"a}dter Stra{\ss}e 39, 1080 Vienna, Austria}
\address[label4]{Department of Mathematics, King's College London,\\ The Strand, London, WC2R 2LS, UK}
\address[label5]{Centro Ricerche Enrico Fermi, Via Panisperna 89 A, 00184 Rome, Italy}

\begin{abstract}
This is a review about financial dependencies which merges efforts in econophysics and financial economics during the last few years.
 We focus on the most relevant contributions to the analysis of asset markets' dependencies, especially correlational studies, which in our opinion are beneficial for researchers in both fields.
  In econophysics, these dependencies can be modeled to describe financial markets as evolving complex networks. In particular we show that a useful way to describe dependencies is by means of information filtering networks that are able to retrieve relevant and  meaningful information in complex financial data sets. In financial economics these dependencies can describe asset comovement and spill-overs. In particular, several models are presented that show how network and factor model approaches are related to modeling of multivariate volatility and asset returns respectively.    
Finally, we sketch out how these studies can inspire future research and how 
they contribute to support researchers in both fields to find a better
 and a stronger common language. 
\end{abstract}


\end{frontmatter}

\newpage
\section{Introduction}\label{sec:intro}

Dependencies in financial markets are complex and can be found at many levels. These markets are influenced by changing economic fundamentals and economic policies around the world. There are also numerous interactions within the asset markets themselves, as well as there is feedback from market participants who have varying expectations about future developments. Many of these processes do not result in measurable streams of data but only manifest themselves in latent variables, like the observable changes in asset prices and other financial data. In order to understand the functioning of financial markets we therefore have to analyze the structures that are present in market data, which in turn means to measure and explain their dependencies. This is of course not a new endeavor, the literature in the field of financial economics has dealt with this for several decades and has developed models that explain the differences in asset returns. Network science has however brought a new approach to the field, as it has enabled us to describe the dependencies of large sets of assets jointly -- and not only factors that explain differences.\footnote{See also \cite{guido_net1} and section \ref{sec:econ}.}

Research on the quantitative aspects of financial markets has always been interdisciplinary, at the intersection of economics, mathematics and statistics. With the advent of networks science however, even more fields became interested in financial markets, especially researchers from physics, computer science and engineering. This has led to many new and important insights into financial markets, but has also made it more difficult to understand the diversity of approaches and to evaluate what the most powerful methods are to study financial dependencies. The overview presented in this review therefore specifically aims at showing different approaches to measuring financial dependencies across disciplines and connecting them to the foundations in financial economics and statistics. This presentation will therefore sometimes be rather wide in scope than deep. This is intentional because we believe that interdisciplinary research is typically only impact-full if it reaches researchers in different disciplines, and when it shows that its authors understand and address research questions across disciplinary borders. We will focus on the analysis of asset returns, in particular the dependencies between stocks, measured by their daily price changes. Most of the presented material will however also apply to any other financial time series in any other frequency.\footnote{See also \cite{tsay} and \cite{dacro} for general introductions to the analysis of financial market data.} We are guided by four questions: how are the returns modeled? How are dependencies measured? How are  filtering or dimensional reduction achieved? How are the measurements mapped into a network representation?

Hence, we first touch upon cornerstones in financial economics and new approaches from econophysics in section \ref{sec:econ}. In section \ref{sec:data} we summarize statistical regularities of asset returns. 
In section \ref{sec:corr} we will explain the basics of calculating correlations and its challenges together with some of its pitfalls. Section \ref{sec:prop} shows the spectrum of correlation matrices and how it is related to finding  stocks with similar behaviors. We then proceed  to present recent approaches from econophysics and financial econometrics. In section \ref{sec:netcorr} we discuss how networks can be derived from correlation measures and we introduce two filtering methods that apply structural constraints on the network. Further we show how clusters and emerging properties can be identified in such networks. 
We then turn to present GARCH models, pairwise estimation methods, and the variance decomposition approach in sections \ref{sec:garch}, \ref{sec:pair} and \ref{sec:var}. In the last section we present some conclusions.

\section{The broader picture of financial market research}\label{sec:econ}
\subsection{Cornerstones in financial economics}

Before discussing recent studies on financial dependencies and networks, it is important to understand the origin of the field, together with its main motives and the scientific language used. 

In financial economics there are two important viewpoints when it comes to asset markets. One is the analysis of market efficiency, the other one is the analysis of the returns of an asset and how it relates to the development of the market and that of other factors. We will focus on the latter. A well-known model in this class is the so-called single index model by \cite{sim}. Here we assume that a vector of returns $r_i$ of an asset $i$ can be described by

\begin{equation}
r_i = \alpha_i + \beta_i r_M + \varepsilon_i \;,
\end{equation}

where $r_M$ is the market return and $\beta_i$ measures how sensitive the stock's return is with respect to the market. $\alpha_i$ is a constant, $\varepsilon_i$ is the residual error. Since we assume that the errors are uncorrelated, i.e. $cov(\varepsilon_i,\varepsilon_j) = 0$ for all $i \neq j$, any correlation between stocks can only come from the market factor $r_M$.

A slight variation of this idea is the capital market pricing model \citep[CAPM,][]{sharpe,lintner}, which includes a risk-free reference interest rate $r_F$, and where the $\beta$ then describes the risk of an investment relative to the market portfolio. Additionally, there is also a large literature that discusses to which extend the $\beta$s are changing over time \citep[see for example][]{vbeta,vbeta2,Radd}.

Variations with more factors have been developed over the years, including the well-known model by \cite{famafrench} which includes factors that account for size and value effects.

\begin{equation}
r_{i} - r_{F} = \alpha_i + \beta_{1,i}(r_M-r_F)+\beta_{2,i} SMB + \beta_{3,i} HML + \varepsilon_i \;.
\end{equation}

Here, $\beta_1$ measures the sensitivity of a stock to market risk. $\beta_2$ and $\beta_3$ measure the sensitivity to the factors SMB (Small Minus Big) and HML (High Minus Low). These factors represent the average return of stocks with certain characteristics in terms of firm size, value, and growth.
Hence, implicitly this model assumes that stocks can be clustered according to these three characteristics.

While much of the differences in asset returns can be attributed to the actions and performance of the firms themselves, there is also ample evidence for the influence of investors. They sometimes evaluate stocks with certain characteristics in a too homogeneous way, disregard differences and thus influence stock prices \citep{barberis,green}. Investors are also not always good in spotting firms that are economically linked, thus missing opportunities to predict returns \citep{cohen}.

It is important to note that the analysis of the comovement on the global level uses more aggregated data \citep[see, e.g.,][]{rigobon_jie}. Most approaches have focused on the analysis of stock market indices or other smaller samples including sectoral indices. A wide range of methods has been applied, among these are unit root and cointegration tests, vector autoregression models, correlation-based tests \citep{contagion,fry}, causality tests \citep{Billio2012}, multivariate GARCH models \citep{DCC}, and models of variance decomposition \citep{yilmazvar}. We will have a closer look at the latter in the following sections.

Another aspect in the literature about financial market dependencies is the question about its determinants \citep{det_inte}. Studies often find that structural similarity of the countries' economies explains only partially the level of comovement of their financial markets. This resulted in a debate about the influence of global sectoral factors \citep{dutt,bekhod,segm_equity}. Previous results hint at an increase in the importance of these factors. \cite{forbchinn} find that cross-country factors and global sectoral factors both are important determinants of stock returns. They also note that changes in global linkages over time might make it difficult to disentangle different influences on asset market comovement. 

\subsection{Statistical physics, econophysics, and complex networks} 

The contribution of statistical physics and the development of econophysics approaches have much to do with the data revolution in social science and economics \citep[see also][]{sandyideas,econohist,cal_humans,stein_adv}. Large-scale investigations on the scaling-properties of asset returns \citep[][see also section \ref{sec:data}]{scaling}  opened the field for many similar investigations of financial data \citep{manstan,jppotbook}. These approaches have rather quickly led to the realization that important challenges in the field are the dependencies of financial data and the complexity of the financial system in general \citep{laloux,plerou}.

Over the course of a decade the analysis of the cross-correlations of asset returns has developed into an analysis of asset markets seen as complex networks, studied with techniques from network theory \citep{man_mst,structure}. 
This has been accompanied by a general influx of network science into the analysis of financial markets in general, fueled of course by the financial crisis that started in 2008 \citep{phys_netw,summer_rev,moglu,bat_complex,ell_cont}.
Further extensions in the methodology are the analyses of bi-partite networks, systems with couplings, and multiple layers of networks \citep{bi_finance,ass_portfolio,multi_net}. 

Describing the details of this evolution is outside the scope of this paper. It however shows that collaboration across disciplines is happening, especially when important problems arise to which established models have too little to offer. It also shows that as so often the connecting element between disciplines is the data, and the need for models that relate it to pressing economic and social challenges.

\section{Preliminaries about asset returns}\label{sec:data}

Dependencies between financial assets are mostly analyzed based on their log daily returns $r_{i,t}$, which are calculated from the stock prices $p_{i,t}$ of asset $i$ at the days $t$ and $t-1$

\begin{equation}
r_{i,t}=\ln p_{i,t} - \ln p_{i,t-1} \; ,
\end{equation}

where $i=1, \hdots, N$ and $t=1, \hdots, T$. $N$ denotes the number of stocks, $T$ denotes the number of observations, in this case these correspond to trading days.

Daily closing prices are the most common data used in the analysis of asset returns. For some applications, e.g. when different markets are involved and trading hours are not perfectly aligned, lower frequencies, for example weekly returns, can be used. Note that the use of logarithmic returns allows to calculate these by summation of the higher frequency returns. On the other end, there is a trend towards the use of higher frequency data in the trading industry that utilizes advances in scalable computing and often employs machine learning.

\begin{figure}
        \begin{center}
        \begin{overpic}[width=\linewidth, trim= 15 180 30 190, clip=true]{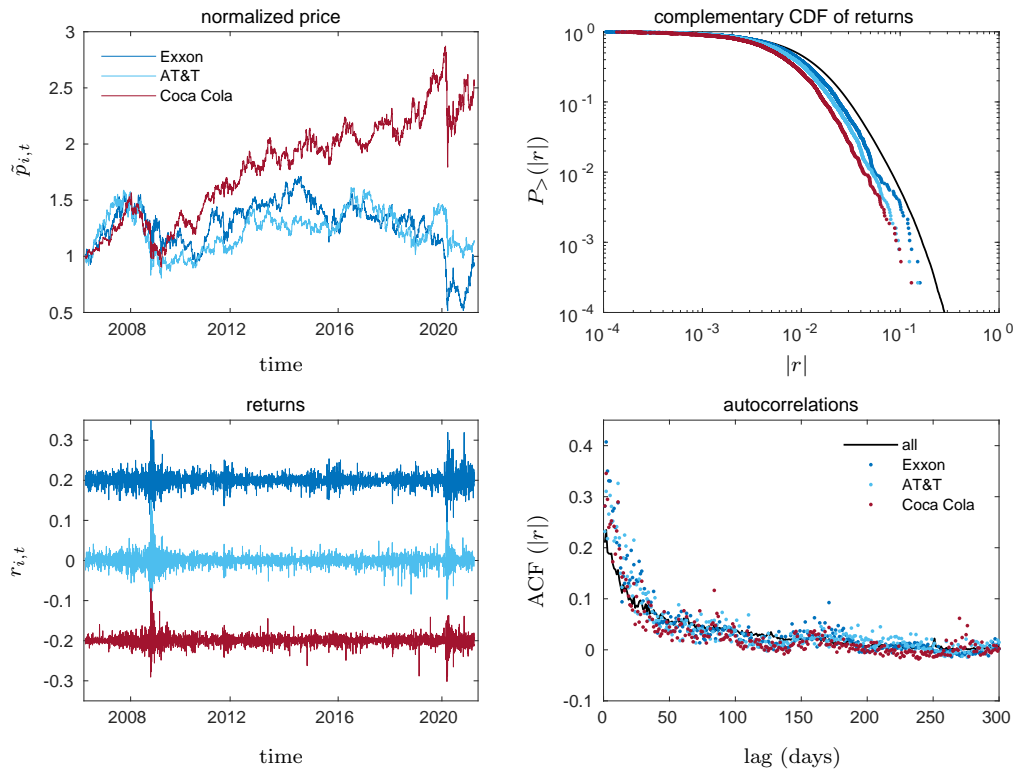}
    \put (26,1) {\scriptsize time}
     \put (73,1) {\scriptsize lag (days)}
       \put (26,39) {\scriptsize time}
     \put (77,39) {\scriptsize $|r|$}
         \put (2,19) {\rotatebox{90}{\scriptsize $r_{i,t}$}}
     \put (2,58) {\rotatebox{90}{\scriptsize $\tilde{p}_{i,t}$}}
     \put (52,16) {\rotatebox{90}{\scriptsize ACF$\; (|r|)$}}
     \put (52,55) {\rotatebox{90}{\scriptsize $P_{>}(|r|)$}}
     \end{overpic}
        \end{center}
        \caption{Daily stock returns and their properties.}{\small The top left panel shows the normalized stock price $\tilde{p}$ for three stocks over time. The bottom left panel shows the corresponding daily returns of these three stocks (two of them plotted with an offset of $\pm 0.2$). The top right panel shows complementary CDFs (CCDFs) of absolute returns on log-log scales for the three stocks, as well as the CCDF for all stocks (black line, with the CCDF $= 1-F(x)$, where $F(x) = Pr[X \leq x]$). The bottom right panel shows the corresponding auto-correlations dependent on the lag in days. }\label{fig:stats}
\end{figure}

For most of the following analyses we use daily data covering the time period from April 2006 until March 2021 ($T = 3,775$ trading days) for $N=404$ stocks that are constituents of the S\&P 500 index. Three exemplary time series of daily prices and returns are plotted in the left panels of figure \ref{fig:stats}. For the purpose of comparison we plot normalized prices $\tilde{p}$, such that

\begin{equation}
 \tilde{p}_{i,t} = \frac{p_{i,t}}{p_{i,1}} \; .
\end{equation}

 While stock prices show generally some upward trend in the long run, we can see that the  returns shown in the bottom left panel fluctuate around zero with some visible spikes and clusters.
 
The volatility of returns can be defined by their sample variance $\hat{\sigma}^2$ as

\begin{equation}
\hat{\sigma}^2 = \frac{1}{T-1} \sum_{t=1}^T \left( r_t - \bar{r} \right)^2 \; .
\end{equation}

One can also define a variance in terms of the expectation  of the time series. A common notation is to state that 

\begin{equation}\label{eq:sigma}
\sigma^2 =  \langle r^2 \rangle - \langle r \rangle ^2   \; ,
\end{equation}

where $ \langle \; \rangle$ denote the expectation. There are also instances where $|r|$ or $r^2$ are used as measures for volatility. This is of course motivated by the fact that $\langle r \rangle$ is typically close to zero and that the intercept of $r$ is in many models estimated separately from the volatility component (more on volatility models in section \ref{sec:garch}).
 
Hence, one aspect of the returns is that their distributions are non-Gaussian, heavily skewed, and fat-tailed.
In fact, the tail of these distributions is well  represented by a power-law, in our case with an exponent of 2.94, as can be seen in the top right panel of figure \ref{fig:stats} \citep[see also][]{scaling}.

Another aspect of the returns is that the volatility is not constant over time. Times with high volatility are often followed by more high volatility. This phenomenon is also called \emph{long memory}  \citep[see also][]{cont}. We can measure this aspect by calculating the autocorrelations, which are the correlations of the time series with lagged versions of itself. Their entirety is also called the autocorrelation function (ACF), which is shown in the bottom right panel of figure \ref{fig:stats}.
The autocorrelations of the absolute returns are typically significantly positive for more than 100 days \citep{dingmemory}. Hence, while we cannot forecast tomorrow's stock prices, some information about tomorrow's volatility is included in today's prices. 

These findings are important for two reasons. First, because they influence how we can analyze the time series of returns and their dependencies. The changing volatility can make it difficult to determine whether two time series are in fact significantly related. Properties of the distribution of stock returns imply that many parametric tests do not apply for this data.
The second aspect of these findings is that they have of course been discussed in streams of literature in different fields and that it is therefore important to understand what the state of the art models are that deal with them -- so that future research on financial networks can benefit from and contribute to them.

\section{Correlations of asset returns}\label{sec:corr}

Pearson's correlation coefficient is the most common measure to describe correlations that are linear. For each pair of stocks $i$ and $j$ it can be calculated as

\begin{equation}\label{eq:C}
C_{i,j}=\frac{\left\langle (r_i- \left\langle r_i \right\rangle ) (r_j- \left\langle r_j \right\rangle ) \right\rangle}{\sigma_i \sigma_j} \; ,
\end{equation}
where  $\sigma_i \sigma_j$ is the product of the standard deviations of the returns of \mbox{stocks $i$} and $j$ (see also equation \ref{eq:sigma}).

Correlation coefficients will always be in the range $[-1,1]$ since we divide by the product of the standard deviations of both variables. An alternative way of measuring dependencies is the covariance, which is given by the numerator of the above expression. This can be useful if one wants to preserve the differences in the scale of the returns for further calculations.

\begin{figure}[tb]
        \begin{center}
        \includegraphics[width=\linewidth, trim= 20 250 20 200, clip=true]{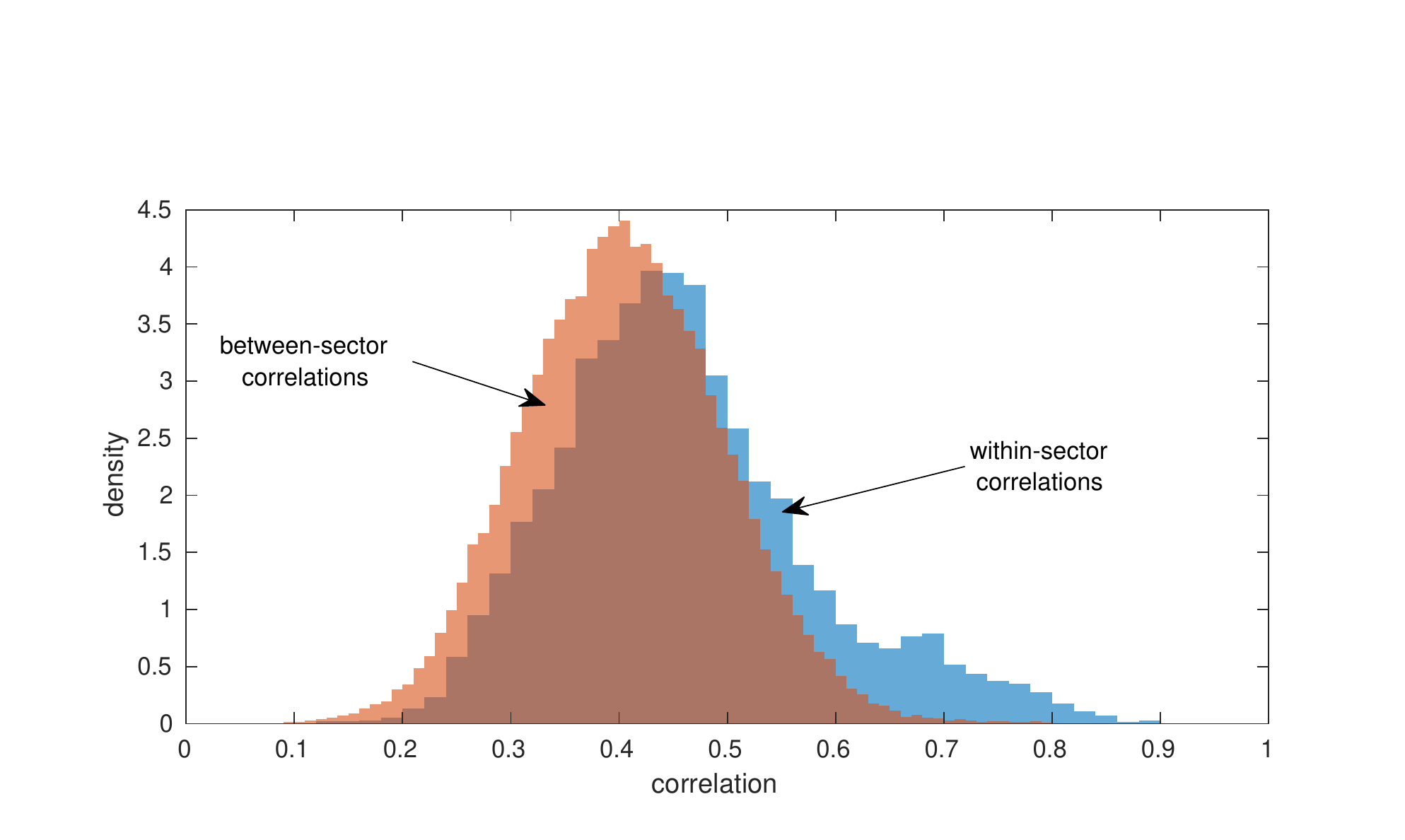}
        \end{center}
        \caption{Histogram of correlation coefficients for the S\&P stocks.}{\small The  two colored areas show the  distributions of correlations between stocks from different sectors (between-sector, orange) and within the sectors (blue). }\label{fig:corrss}
\end{figure}

Figure \ref{fig:corrss} shows distributions of correlation coefficients for the S\&P sample of stocks. While the distribution of the correlation coefficients of stocks from different sectors (between-sector) shows similarity to a Gaussian distribution, the distribution of within-sector correlations has a higher mean and shows skewness.

\begin{figure}[p]
        \begin{center}
        \begin{overpic}[width=\linewidth, trim= 40 120 40 150, clip=true]{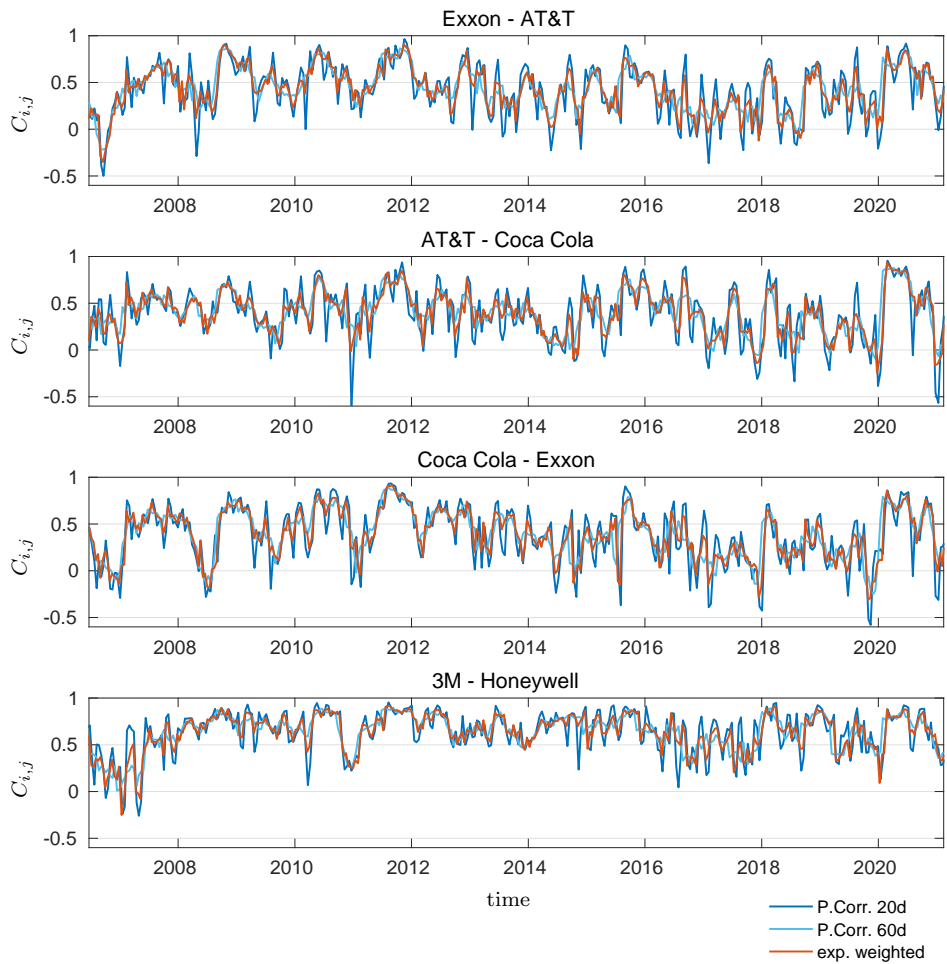}
         \put (3,83){\rotatebox{90}{\scriptsize $C_{i,j}$}}
         \put (3,61.5){\rotatebox{90}{\scriptsize $C_{i,j}$}}
         \put (3,40){\rotatebox{90}{\scriptsize $C_{i,j}$}}
         \put (3,18.5){\rotatebox{90}{\scriptsize $C_{i,j}$}}
     \put (49,8){\scriptsize time}
     \end{overpic}
        \end{center}
        \caption{Stock correlations with different window lengths.}{\small These panels show four examples for the development of correlations over time. In each of them the dark blue line shows Pearson's correlations calculated in a window with $\Delta t = 20$ days. The dark blue line shows these correlations calculated  for $\Delta t = 60$ days. The red line shows exponentially weighted correlations with a time window of $\Delta t=60$ days (and $\theta=20$). The top three panels show examples where the stocks come from different sectors. The bottom panel shows an example of within-sector correlations.}\label{fig:corr1}
\end{figure}

In many applications we are interested in a dynamical analysis of these correlations. Pearson's correlations coefficients are therefore often calculated by applying a running window approach, as illustrated in figure \ref{fig:corr1}. Equation \ref{eq:C} is in this case applied only to the returns inside a time window with length $\Delta t$. Here the first three panels show pairs of stocks from different sectors, while the last panel shows an example for a pair of stocks which both belong to \emph{Non-cyclical consumer goods}. 

Using running windows can however lead to spurious results when the window size $\Delta t$ is too small. Correlations are then likely to be exaggerated. Also, the comparison of these correlations is difficult since the normalization by the changing variance in equation \ref{eq:C} will drive the results.

\begin{figure}[htb]
        \begin{center}
        \begin{overpic}[width=\linewidth, trim= 25 275 25 295, clip=true]{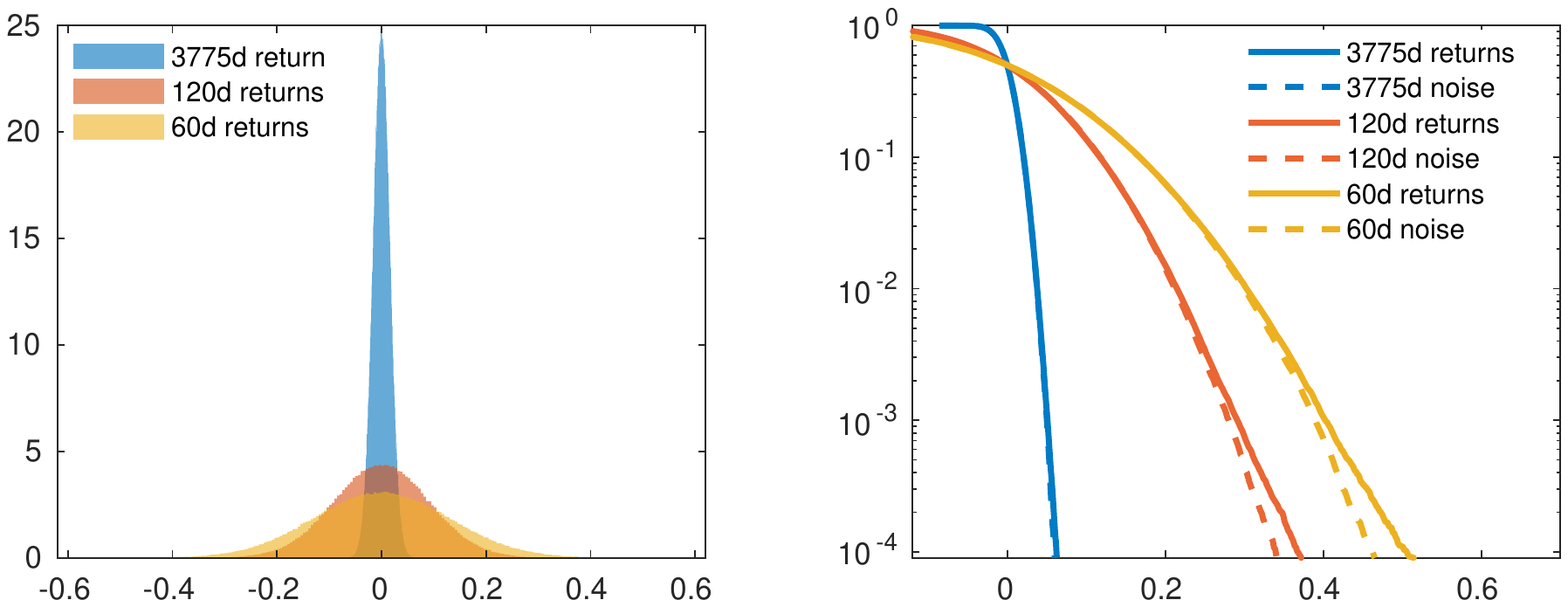}
         \put (1,18){\rotatebox{90}{\scriptsize density}}
          \put (50,18){\rotatebox{90}{\scriptsize $P_>(C_{i,j})$}}
           \put (22,1){\scriptsize correlation}
           \put (72,1){\scriptsize correlation}
      	\end{overpic}
        \end{center}
        \caption{Correlation coefficients from permutated returns and from noise.}{\small The left panel shows  probability distribution functions that were obtained by calculating the correlation coefficients of randomly drawn pairs of permutated time series with different length from the S\&P returns data ($\Delta t=$ 60, 120, 3775 days). The right panel shows the comparison of the tails of these distributions (as a CCDF) with corresponding time series composed of independent and identically distributed noise.}\label{fig:sign}
\end{figure}

The assessment of the significance of correlations therefore deserves a closer look. It is well known that both the error of the correlation coefficient itself as well as intervals for significance levels that determine if a correlation coefficient is significantly different from zero can easily be calculated making use of the corresponding t-distribution. The latter is achieved by a parametric test that essentially answers the question what range of correlation coefficients can be expected from  time series with a specific length when they are uncorrelated and normally distributed (i.e., when they consist of random variables that are independent and identically distributed). In order to evaluate in how far the violation of this assumption affects the correlations, we can compare
correlation coefficients of permutated time series pairs of returns (of which the expected correlation is zero) with that of uncorrelated noise \citep[see also][]{aste_corr}. Figure \ref{fig:sign} shows the distributions of correlation coefficients, both for permutated real data as well as for independent normally distributed noise (we draw 500,000 samples for $\Delta t = 20$ and $60$ days with random starting points, and we calculate all possible correlation coefficients for $\Delta t=3775$, which corresponds to the entire sample).

First of all, it is important to realize that the significance of correlation coefficients becomes much more difficult to asses once the length of the time window becomes smaller. We can clearly see this by comparing the distribution of obtained correlation coefficients for the entire sample period with that associated with the time window length used before, $\Delta t =60$ days (left panel). With regards to our results on the dynamics of correlations presented in figure \ref{fig:corr1} this means that the correlations between most stocks are only significantly positive for some time windows.

A second aspect is the slight inaccuracy when parametric tests are applied to this correlation measure of asset returns. In order to visualize the effect it is necessary to focus on the tails of the distributions (right panel in figure \ref{fig:sign}). When we compare these tails with the corresponding distributions of independent normally distributed noise, we observe slightly too many extremes in the correlation coefficients of the permutated returns, especially for small $\Delta t$.
Hence, relying on a parametric test for the correlation coefficients can in this case lead to an overstatement of their significance \citep[see also][]{r_non}.

It is therefore of most importance to select time windows such that true dependencies can be reliably identified as significant. In some applications the choice of the window length can be a trade-off, especially when there is the necessity to associate networks with specific economic events.
It turns out that an improvement can be achieved by introducing a weighting of the observations within a time window. By putting high weights on the latest returns and successively lower weights on past observations, the correlations can be smoothed while retaining a relatively high time resolution \citep{wcorr}. In this case a weighting applies to the sample mean, the variance and the covariances in equation \ref{eq:C}. The weights $\sum_{t=1}^{\Delta t} w_t = 1$ follow
\begin{equation}
w_t = w_0 \; exp \left( \frac{t-\Delta t}{\theta} \right) \;, \;t \in \{1,2,\dots, \Delta t\}   \; ,
\end{equation}
which describes an exponential weighting scheme.\footnote{$\theta$ is the weights' characteristic time, with $\theta > 0$. $w_0$ is a constant, in the case of Pearson's correlations equal to $(1 - exp(-1 / \theta)) / (1 - exp(-\Delta t / \theta))$.}

It should be noted that several other techniques exist to work around the unreliability of Pearson's correlations or that try to filter the returns for certain effects. 

One technique for such a filtering is the use of partial correlations, which investigates how the correlation between two stocks  $i$ and $j$ depends on the correlation of each of the stocks with a third variable, typically a stock index $m$ as the mediating variable.
The partial correlation $\rho_{i,j|m}$ can be viewed as the residual correlation between stocks $i$ and $j$, after subtraction of the contribution of the correlation between each of the stocks with the index.

Some studies have extended this approach to the analysis of the ratio of partial and raw correlations and the metacorrelations between different markets \citep{ICF,ICF1,globalstock}.

Another alternative is the use of correlation measures that are non-para\-metric, like rank-correlation, or the tail dependence. The latter is a non-parametric estimator of the tail copula that provides a measure of dependence focused on
extreme events \citep{c_measure}. Also a transformation of the returns can improve measurements if it brings their distribution closer to that assumed by a particular model or test.

\section{Properties of correlation matrices}\label{sec:prop}

In the previous sections we have introduced the correlation matrix of returns and have given examples for the empirical results that we typically obtain by calculating correlations coefficients from stock returns.

In this section we want to connect these findings to the most important theoretical concepts about the decomposition of correlation matrices, since these are useful foundations to understand the structure contained in stock markets. 

In equation \ref{eq:C} we have already shown how to calculate correlation coefficients. An equivalent statement is the definition of the sample \mbox{correlation $R_{i,j}$}.

\begin{align}
R_{i,j} = \frac{ \sum_{t=1}^T \, (r_{i,t} - \bar{r_i}) \, (r_{j,t} - \bar{r_j})}
{\sqrt{ \sum_{t=1}^T \, (r_{i,t} - \bar{r_i})^2} \sqrt{ \sum_{t=1}^T \, (r_{j,t} - \bar{r_j})^2}} \; .
\end{align}

Note that we use  sum notation here instead of the brackets as in equation \ref{eq:C}, the mean of $r$ is thus denoted by $\bar{r}$. It is worth to be familiar with both notations since they are both frequently used in the literature.

We also note that some contributions to the literature, especially about the eigenvalue spectrum, assume standardized returns $\tilde{r}$. In this case the sample correlation $\tilde{R}$ can be written as follows and $ R = \tilde{R}$

\begin{align}
\tilde{R}_{i,j} = \frac{1}{T} \sum_{t=1}^T \, \tilde{r}_{i,t} \, \tilde{r}_{j,t} \; .
\end{align}

We can use the spectral analysis of the correlation matrix to infer a few properties with respect to the level of noise versus structure that it contains. 
A prominent method concerned with the spectrum of correlation matrices of uncorrelated data is Random Matrix Theory (RMT). Assuming time series of white noise with zero mean and unit variance, \cite{MP} state that the distribution of eigenvalues $\lambda$ of its correlation matrix follows

\begin{figure}[htb]
        \begin{center}
        \includegraphics[width=\linewidth, trim= 45 275 50 270, clip=true]{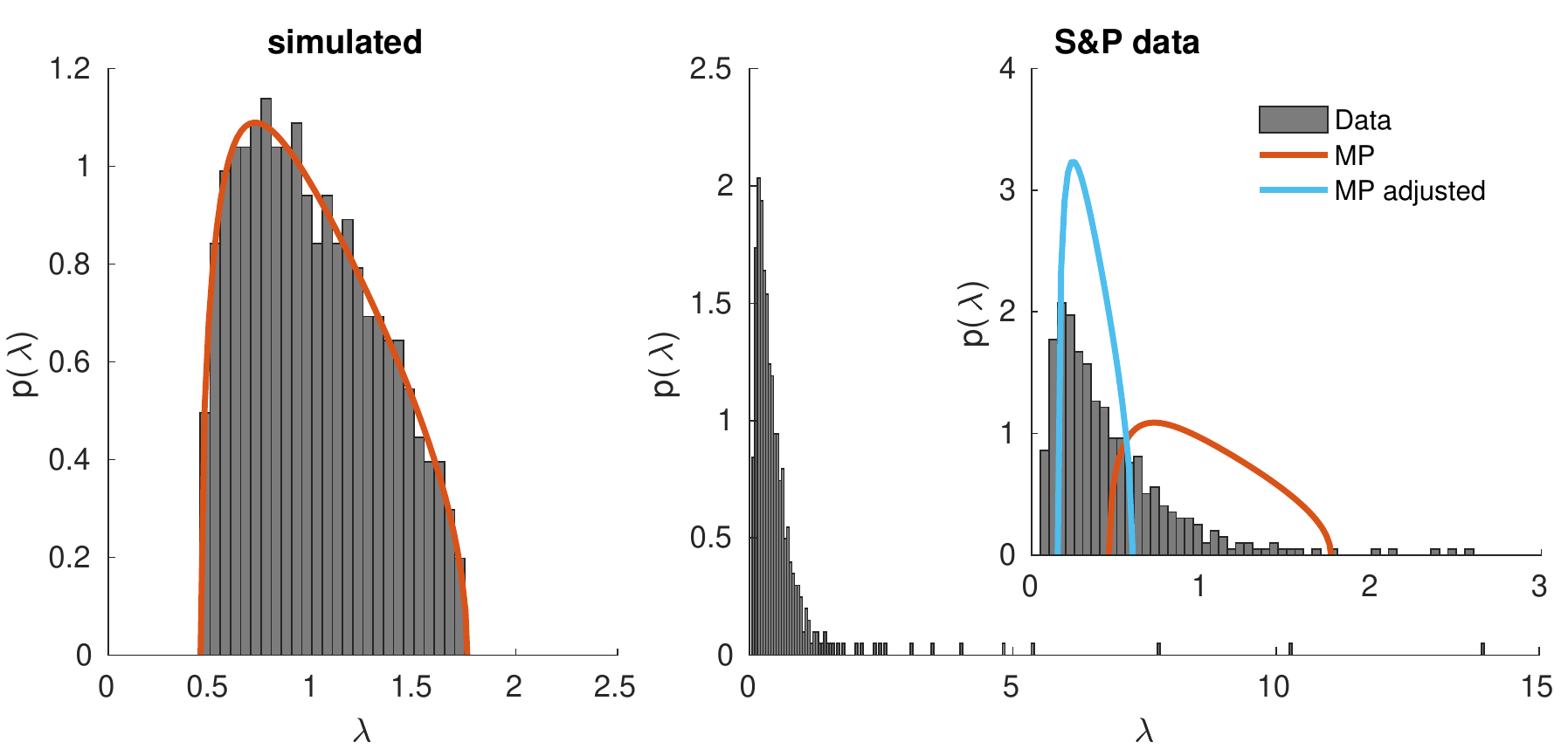}
        \end{center}
        \caption{Histograms of eigenvalues of correlations matrices and MP-spectrum.}{\small The left panel shows the histogram of the eigenvalues for random time series with the same $Q$ as the real data. The MP spectrum fits the data exactly. The right panel shows the distribution of eigenvalues of the market data. The largest eigenvalue, $ \lambda_1 = 170 $, is not shown. The insert shows the eigenvalues in the range predicted by MP. Large parts of the market's spectrum deviate from MP. When one adjusts the MP spectrum for the leading eigenvalue by setting $\sigma_r = 0.58$ some qualitative alignment can be achieved.}\label{fig:rmt}
\end{figure}

\begin{equation}
p(\lambda) = \frac{Q}{2 \pi \sigma_r^2} \frac{\sqrt{(\lambda_+ - \lambda)(\lambda - \lambda_-)}}{\lambda} \; ,
\end{equation}\label{eq:rmt}

where $Q = T/N$ adjusts for the ratio of the length of the time series versus the sample size. $\sigma_r$ is a further adjustment parameter which is 1 in the case of standardized data. It can also be used to compensate for single eigenvalues outside the predicted spectrum. 

The boundaries of the spectrum are given by
\begin{equation}
\lambda_\pm = \sigma_r^2 \left( 1 + \frac{1}{Q} \pm 2 \sqrt{\frac{1}{Q}} \right) \; .
\end{equation}

If $N \to \infty$ and $T \gg N$ the eigenvalues should all lie in the interval $[\lambda_-,\lambda_+]$ \citep{laloux,plerou,breview}.

An example is shown in figure \ref{fig:rmt}. Artificial time series, i.e., simulated returns drawn from a normal distribution that are uncorrelated, produce a histogram of eigenvalues that agrees with the prediction from equation \ref{eq:rmt} (left panel). 

We should clarify that this analysis does not suggest that the MP prediction describes the spectrum of eigenvalues from empirical asset return data perfectly \citep[see also][]{mp_details}. It merely shows that the number of eigenvalues that are far outside the predicted spectrum is rather small. 

It is very common to find one large eigenvalue, which is associated with the market mode.  Its corresponding eigenvectors can for example be used to calculate the weights of stocks in a market index or conversely for the construction of certain optimized portfolios. The other eigenvalues outside the MP-spectrum generally describe groups of stock that show a systematic behavior that is different from the market.
It should therefore be possible to generate good approximations of the correlation matrix by models that decompose it using a number of factors $k = 1 \hdots K$ much smaller than $N$. We will now introduce two approaches for such a decomposition.

The correlation matrix defined above is by construction an $N \times N$ symmetric matrix that can be diagonalized. This
is the basis of the well known Principal Component Analysis (PCA). In fact, for many cases it makes sense to go one step further and to use this decomposition to model the returns themselves.  We therefore decompose changes in $r_{i,t}$ into  decorrelated contributions of decreasing variance. In terms of the eigenvalues 
$\lambda_k$ and eigenvectors $v_k$, the decomposition reads
\begin{equation}\label{eq:decomp}
r_{i,t} = \sum_{k=1}^N \sqrt{\lambda_k} v_{k,i} \, \varepsilon_{k,t} \; ,
\end{equation}

where $v_{k,i}$ denotes element $i$ in the $k$th eigenvector and $\varepsilon_{k,t}$ are uncorrelated  random variables with unit variance.

We note that the formulation above is very common in any physics textbook, yet statistics textbooks in economics  or business often use a different viewpoint, which we will also briefly sketch out. Let us assume that our sample of returns (our variables) are stored in a matrix $X$ with dimensions $N \times T$. We can then calculate a new set of variables, 
 the principal components $F$, which are linear
combinations of the original variables (for examples for a dimensional reduction of the system).
Hence, we search for
$F=Y X$. The column vectors in
$Y$ carry the weights for each new variable, and it can be
shown that the solution to this problem amounts to solving for

\begin{equation}\label{eq:factor}
F=V X \; ,
\end{equation}

where $Y = V$ contains the eigenvectors of the correlation
matrix $R$ \citep[typically ordered by descending eigenvalues, see][]{Jobson}. 
The eigenvectors with the highest eigenvalues
account for a large amount of the variance in the data, while low
eigenvalues stand for eigenvectors and components that contribute very
little to the variance and can therefore be neglected to describe $X$. 

The difference between these two approaches is what is explained on the left hand side (note that we can derive the factor loadings by scaling the eigenvectors with the variance $\sqrt{\lambda}$). Equation \ref{eq:decomp} states the decomposition of the returns as a stochastic model, equation \ref{eq:factor} states an identity that describes the factors of said decomposition. For a more detailed example the reader might refer to \cite{anshul}.

\section{Correlations and Networks}\label{sec:netcorr}

\subsection{Networks, filtering, and dimensional reduction}\label{sec:reduct}

Financial markets are systems with a large number of elements, in our case the number of stocks $N$. All these stocks can be differently affected by one another. We aim to infer the network of most relevant links by studying the mutual dependencies
between stocks based on an analysis of correlation measures.
 In other words, we search for stocks that behave similarly and we want to link
them with edges in the network. Conversely, we do not want to directly
connect stocks that behave independently. In many cases it is further necessary to determine which nodes can be clustered together. This information can either be used to group these stocks, it can also be used to dimensionally reduce the system and to define a new set of nodes based on these clusters' aggregations.  

Hence, there are typically two considerations to be dealt with when it comes to mapping stocks into networks. The first one is which links to include into a network given the full set of stocks. The second one is the dimensional reduction on the part of the nodes (stocks) itself. The later becomes more important once the size of the network becomes large and can no longer be usefully visualized, or when subsequent analyses of the network are not feasible for the entire sample. The problem of a dimensional reduction can either come into play at the stage of the visualization of the network, it may also come into play earlier, namely at the stage of estimating correlations within part of an econometric model. 

Let us start with the simplest case, which is the mapping of the correlation matrix into an adjacency matrix by using a threshold rule. The adjacency matrix is a matrix where rows and columns are labeled according to the nodes (stocks) in the network and where the entries in the matrix describe whether certain nodes are connected (adjacent) to each other.
In this case we create an adjacency matrix $A$ with dimensions $ N \times N$, identical to our correlation matrix. One constructs this adjacency matrix by setting a threshold for the correlation coefficient and defining that the elements $A_{i,j}$ are 1 if the corresponding correlation coefficients are above a certain threshold and zero otherwise. The diagonal elements of $A$ are typically zero, unless the concrete example requires to express a self-loop. The result will be an undirected unweighted network. Weighted networks can be obtained by using the absolute value of the correlation coefficients (or transformations of these) as weights of the edges. Examples are shown in appendix A.\footnote{The network visualization in this review were produced in Gephi \citep{gephi}, mostly using the algorithm by \cite{hu}.}

While these networks can be informative there are also some drawbacks. Their visualizations can sometimes change drastically depending on the threshold level that is chosen. They often contain a very high number of edges and it can be difficult to asses the significance of these connections. Resulting networks may also contain several components. In other words -- they do not necessarily provide a meaningful and robust simplification of the system by means of network topology.\footnote{It should be noted that on an only slightly generalized level the problem of finding the `relevant' links in a network connects to an incredibly large amount of literature in mathematics, computer science and statistics, all in some way concerned with finding sparse versions of covariance matrices. A noteworthy example, outside the scope of this review, are for example specific convex optimization (Bayesian) methods also known as the Lasso \citep[see, for example,][]{lasso1,lasso2}.}

The second consideration in creating networks is the dimensional reduction of the system itself. We will illustrate this by applying a simple clustering technique to our data set. The output of such a clustering can be used to reduce the number of nodes in the system.
Most clustering techniques will require to asses the similarity of stock returns by some distance measures. The correlation matrix contains information about similarity, yet it makes sense to transform it in such a way that it reflects a metric. Hence, the following transformation can be used to derive a distance matrix $D$ \citep[see also][]{man_mst}

\begin{equation}\label{eq:dist}
D_{i,j} = \sqrt{2 (1-C_{i,j})} \; .
\end{equation}

Hence, the elements of the correlation matrix are transformed such a way that $D$ has a lower bound of 0 that can only be obtained by perfectly correlated stock returns.

One can then use the matrix $D$ in an algorithm that joins the stocks together based on their distances. Here we will focus on a hierarchical approach.  Assume that at the start all stocks are in a group (or cluster) on their own. We will start by joining the pair of stocks with the lowest distance. Before searching for the next shortest distance between all groups, we have to make a choice about how to define the distance between the now joined first cluster and the remainder of the stocks. Common solutions are to use the average or the minimum. Hence, by repeatedly joining the groups that are closest together and by updating our distance matrix for the groups, we will ultimately connect all stocks in a hierarchical structure -- at different levels of the distance measure. 

\begin{figure}[tb]
        \begin{center}
        \begin{overpic}[width=\linewidth, trim= 60 0 140 0, clip=true]{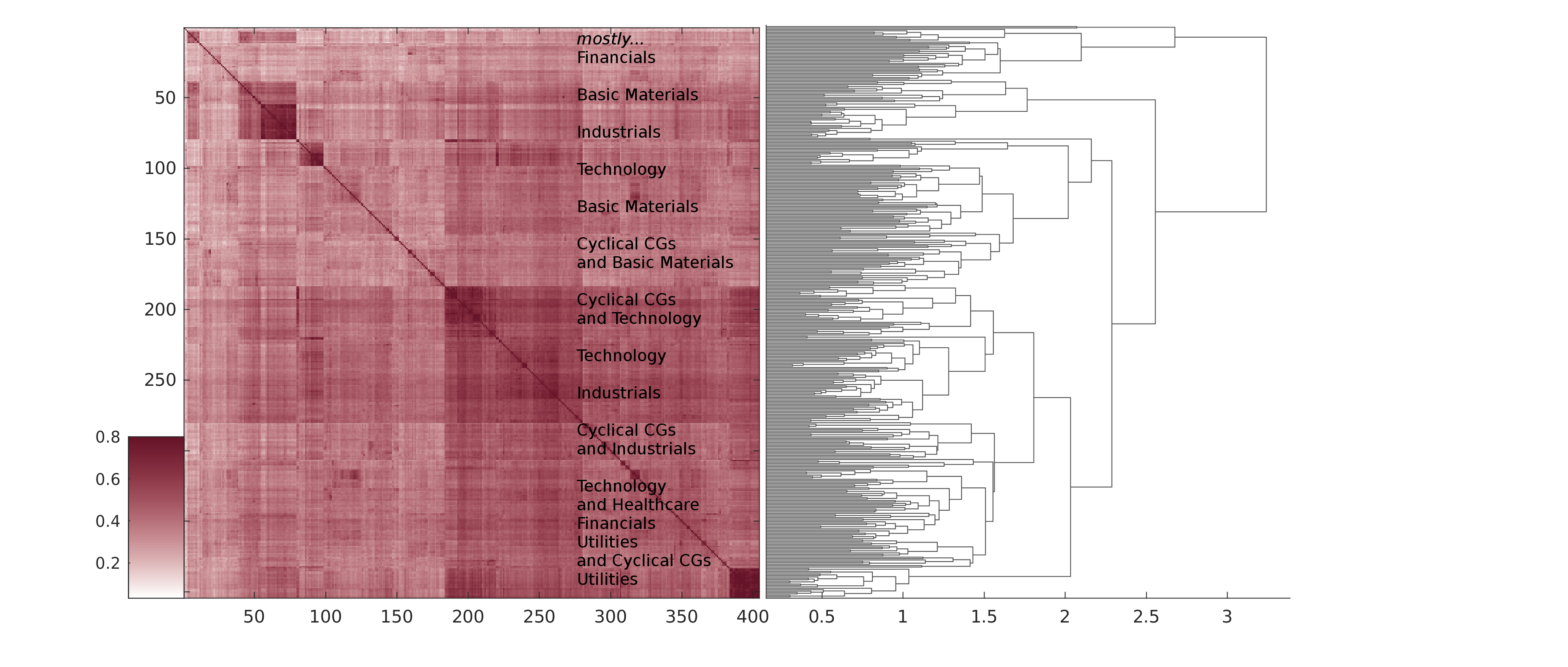}
        \put (3,7){\rotatebox{90}{\scriptsize correlation}}
        \put (28,1){\scriptsize stocks}
         \put (71,1){\scriptsize D -- distance}
         \put (15,55){\scriptsize \textbf{ordered correlation matrix}}
         \put (70,55){\scriptsize \textbf{dendrogram}}
        \end{overpic}
        \end{center}
        \caption{Color-coded correlation matrix and dendrogram of the hierarchical clustering for the constituents of the S\&P 500.}{\small The rows and columns in the left panel are ordered according to the dendrogram in the right panel. Sector classifications for the clusters are superimposed based on the most frequent label within a cluster.}\label{fig:dend}
\end{figure}
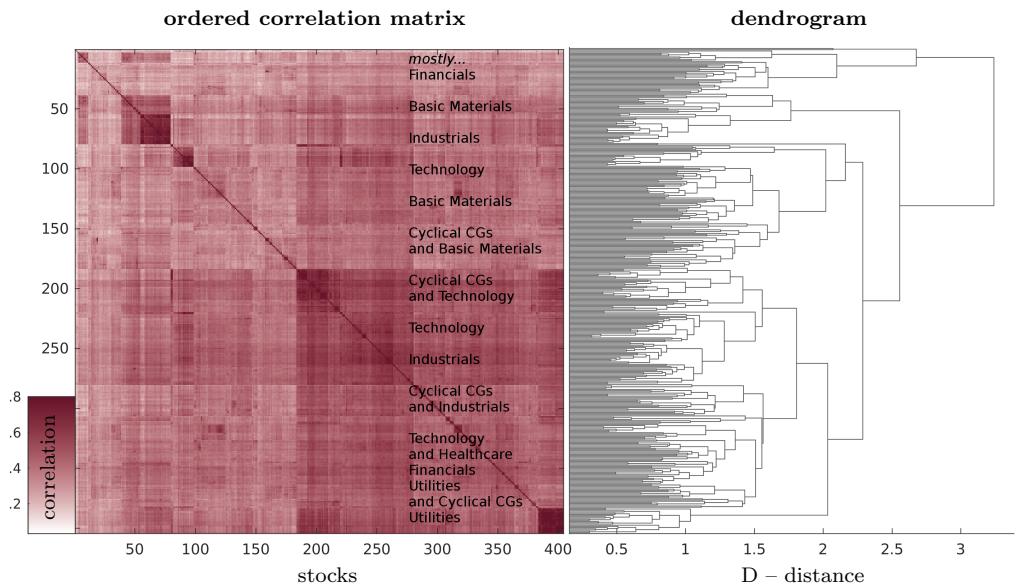

This result can be demonstrated by the dendrogram shown in the right panel of figure \ref{fig:dend}. It illustrates which stocks would be sorted into the same group given a certain threshold for the distance. The left panel shows a visualization of the correlation matrix where the rows and columns are sorted according to the dendrogram. The regularities in the color-coding hint at uncovered structures in the correlation matrix. 

While such a grouping can certainly be visualized as a network, we should note that a few more steps would be needed to obtain proper adjacency matrices. First, one would have to set a threshold for a cut-off point in the dendrogram. Second, one would have to set a rule that defines how connections within the identified clusters are dealt with, or if new nodes, based on the identified clusters, would be constructed.

The number of clustering techniques is very large, as are its applications to networks, which leads to the problem of community detection \citep[see also][]{Jobson,NewBook}. Studies on the use of correlations in this context include \cite{com_mahon} and \cite{conf_masuda}.  An excellent comparison of community detection algorithms can be found in \cite{cd_comparison}.

The literature in financial economics has a slightly different approach on the filtering of dependencies. Since the sample sizes discussed in this strand of the literature are typically slightly smaller than in econophysics, many approaches aim at deriving full matrices of weighted dependency measures. The problems that these approaches face with respect to system size are however tightly related to the problems that we have just sketched out. Dimensional reductions are in these models often made in the process of estimating the dependencies in order to circumvent numerical limitations. We will come back to some of these issues in sections \ref{sec:garch} to \ref{sec:var}.

In the following we will now present approaches that provide an efficient filtering of the system of dependencies and we show how these can be extended to an endogenous clustering.\footnote{We note that further important contributions to the research on correlation-based networks include \cite{structure}, \cite{portfnet} and \cite{corrhier}.}

\subsection{Minimum spanning tree}\label{sec:mst}

The minimum spanning tree (MST) is a tool to analyze and filter the information contained in the correlation structure of a set of financial assets. We look at the correlation matrix as the adjacency matrix of a network and generate an MST
on this network in order to retain the most significant links.
The complete network is represented by links which weights are determined by the corresponding correlation coefficients. This complete network has $\frac{N(N-1)}{2}$ links, representing the full information of the system.

The MST approach by itself is a rather old mathematical problem that has been discussed since the 1930s \citep[and solved, see][]{segbook}. A minimally spanning tree (MST)  consists of the subset of edges (links) of a graph that connects all the vertices (nodes) using the minimal total edge weight. Hence, it can be interpreted as the ``least expensive'' connected graph. It contains no loops and the number of edges is $N-1$, given that $N$ is the number of vertices (nodes).\footnote{MST being a classical math problem, you will find the terminology of graphs, vertices and edges  in the related literature instead of networks, nodes and links. The former terms are common in settings when abstract networks are discussed, yet by today they are often used interchangeably.}

The details of applying MSTs to asset markets were for the first time presented by \cite{man_mst}.
In order to describe the dependencies of stock returns as a MST we first have to define distances between our stocks, and we do this again by applying equation \ref{eq:dist}.  Several algorithms for the construction of MSTs have been proposed, here we are using hierarchical clustering and single linkage \citep{mst_sl,kruskal,prim}.

To construct the MST, let us assume that at the start our network consists of all the stocks (nodes) in our data set, yet no links exist between them. We start by sorting all the pairs of stocks by their distances in ascending order. We start by connecting the stocks with the shortest distance. We then proceed with all the remaining pairs in the order of their distances while omitting links that would create a loop. 

It can be shown that this algorithm does in fact produce the MST with the lowest overall edge-weight (sum of distances) possible. The proof goes along the lines of showing that at each step of the algorithm we are using the lowest weight edge to connect the single parts of the so-called forest of clusters of nodes into a single tree, hence, there can be no improvement to the total edge weight by swapping edges.

\begin{figure}[p] 
        \begin{center}
        \begin{overpic}[width=0.9\linewidth, trim= 10 60 10 60, clip=true]{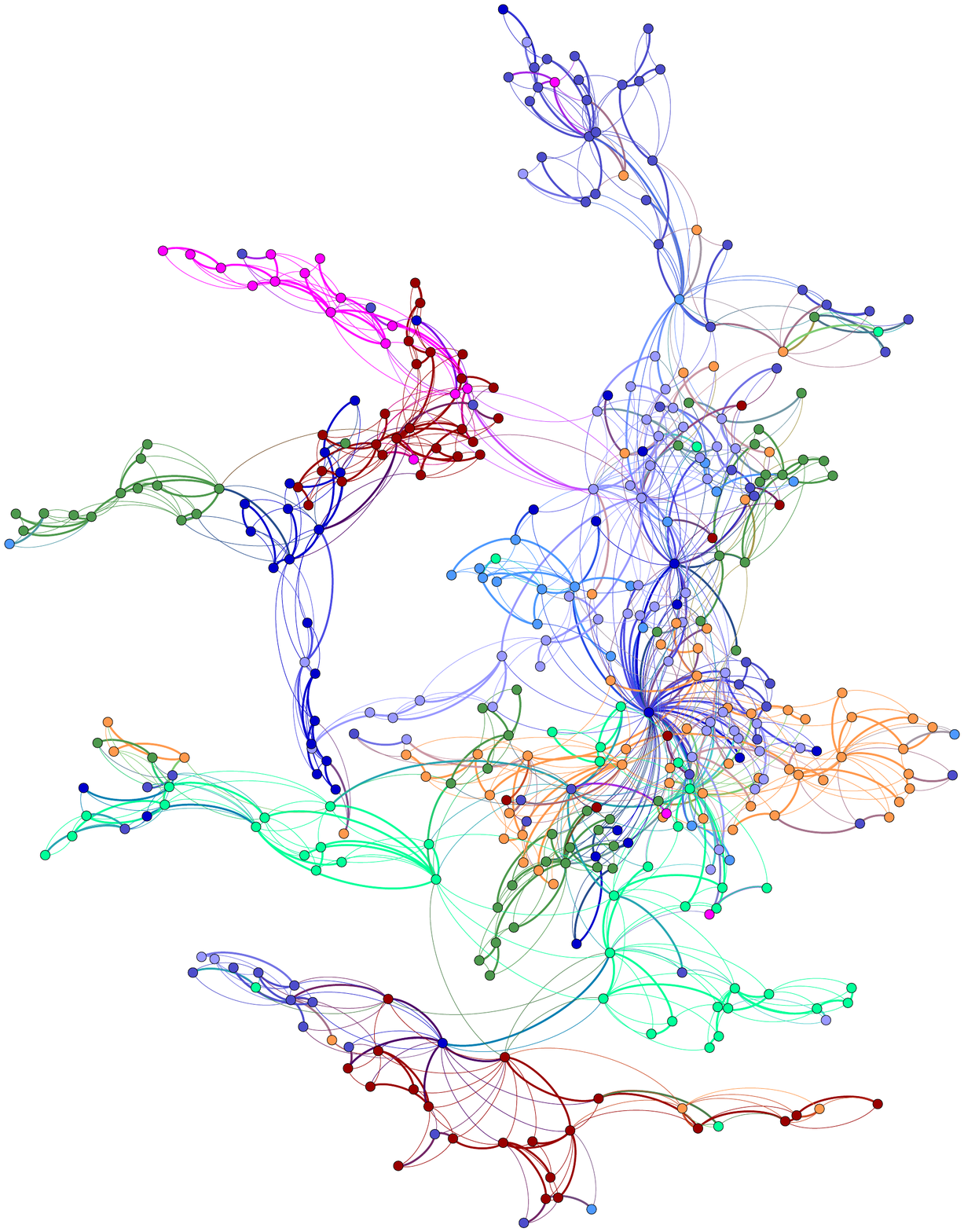}
   \put (50,43) {\footnotesize \textbf{\contour{white}{Honeywell}}}
    \put (35,69) {\footnotesize \textbf{\contour{white}{Exxon}}}
      \put (54,55) {\footnotesize \textbf{\contour{white}{3M}}}
      \put (12,39) {\footnotesize \textbf{\contour{white}{AT\&T}}}
      \put (21,57) {\footnotesize \textbf{\contour{white}{Coca Cola}}}
  \end{overpic}
        \end{center}
        \caption{PMFG for the constituents of the S\&P 500.}{\footnotesize Connections of the corresponding MST are shown by bold edges. Color-coding by sector classification: pink - Energy, bright blue - Basic Materials, bright purple - Industrials, dark purple - Cyclical Consumer Goods, dark blue - Non-cyclical Consumer Goods, bright green - Financials, dark green - Healthcare, orange - Technology, brown - Utilities. This visualization highlights the clustering features of the PMFG and not its planarity -- thus some edges cross. }\label{fig:pmfg}
\end{figure}

This algorithm will in the early stage typically produce several components of connected stocks that will gradually grow together to form a connected network. Often one will find several star-like regions that are made up of stocks that belong to similar sectors of the economy. A visualization is provided as part of figure \ref{fig:pmfg}. The MST algorithm provides a hierarchical classification of the stocks \citep[see also][]{risk1}. For a further analysis of asset trees see also \cite{onnela2}.

\subsection{Planar maximally filtered graphs}\label{sec:pmfg}
\subsubsection{Construction principles}

The MST does by its very nature lead to a relatively drastic reduction in information by describing the dependencies by only those links that are absolutely necessary to produce a connected network. It is therefore natural to ask if further algorithms can be formulated, which allow for more links and thus information to be represented in more detail.

A side-effect of the construction of the MST is that it produces networks that can without much effort be displayed in a 2-dimensional plain representation. When we want to increase the amount of information in the network by adding more links we therefore have to define a consistent rule that stops us from adding links that make the network become too complex. The most cited method for this is the planar maximally filtered graph (PMFG) described in \cite{pmfg1} and \cite{pmfg}. The idea here is to include all links that
produce a graph that is planar.
This corresponds to creating a network embedded on a sphere with genus equal to 0. In principal this method can be applied to surfaces of different genera, yet in practice the most useful one is that of the planar case \citep{tpmfg2}.

To construct the PMFG we again define distances according to equation \ref{eq:dist} and employ an algorithm similar to that used for the construction of the MST. Hence, we add links in ascending order based on the distance measure. The difference is that now we do not discard links between already connected stocks right away. Before adding links we confirm that their addition does not produce a non-planar graph.

Checking this feature is unfortunately not entirely trivial. The problem of proving that a graph is planar goes back to work done by Leonard Euler. It turns out that it is relatively easy to show which conditions are necessary for planarity and non-planarity, but that sufficient conditions are relatively tricky. The problem was however solved by \cite{kuratowski} who showed that a graph is non-planar if and only if it contains a subgraph homeomorphic to two particular graphs that are mostly described as $K_{3,3}$ and $K_5$.\footnote{Two graphs are homeomorphic if one is a subdivision of another, or they are both subdivisions of some third graph. $K_{3,3}$ is a bipartite structure with 3 nodes in each set, $K_5$ in a completely connected set of 5 nodes.}

We will not go further into the details of this proof, the problem is however rather apparent when one sketches a network with 4 nodes and connects them all with a pen on a piece of paper. Adding another node and connecting it to all other nodes is impossible. Hence, cliques with more than 4 nodes are not possible in a PMFG, and the number of links in a PMFG is always equal to $3(N-2)$.

A visualization of a PMFG is shown in figure \ref{fig:pmfg}. We have labeled the positions of stocks that were mentioned before. \emph{Honeywell} appears as a root in this network, as it comoves closely with many other stocks. Finding one stock as a dominant hub in the PMFG is a common finding in PMFGs of asset markets. The structure of the periphery of the network partly coincides with the sector classifications, while the center of the network is less segmented. For more detailed applications of the PMFG the reader may refer to \cite{pmfga2} and \cite{pmfga1}. A computationally more efficient algorithm suitable for large data sets is developed in \cite{tpmfg}.

\subsubsection{Clustering}\label{sec:dbht}

\begin{figure}[p]
        \begin{center}
        \begin{overpic}[width=0.95\linewidth, trim= 0 150 0 160, clip=true]{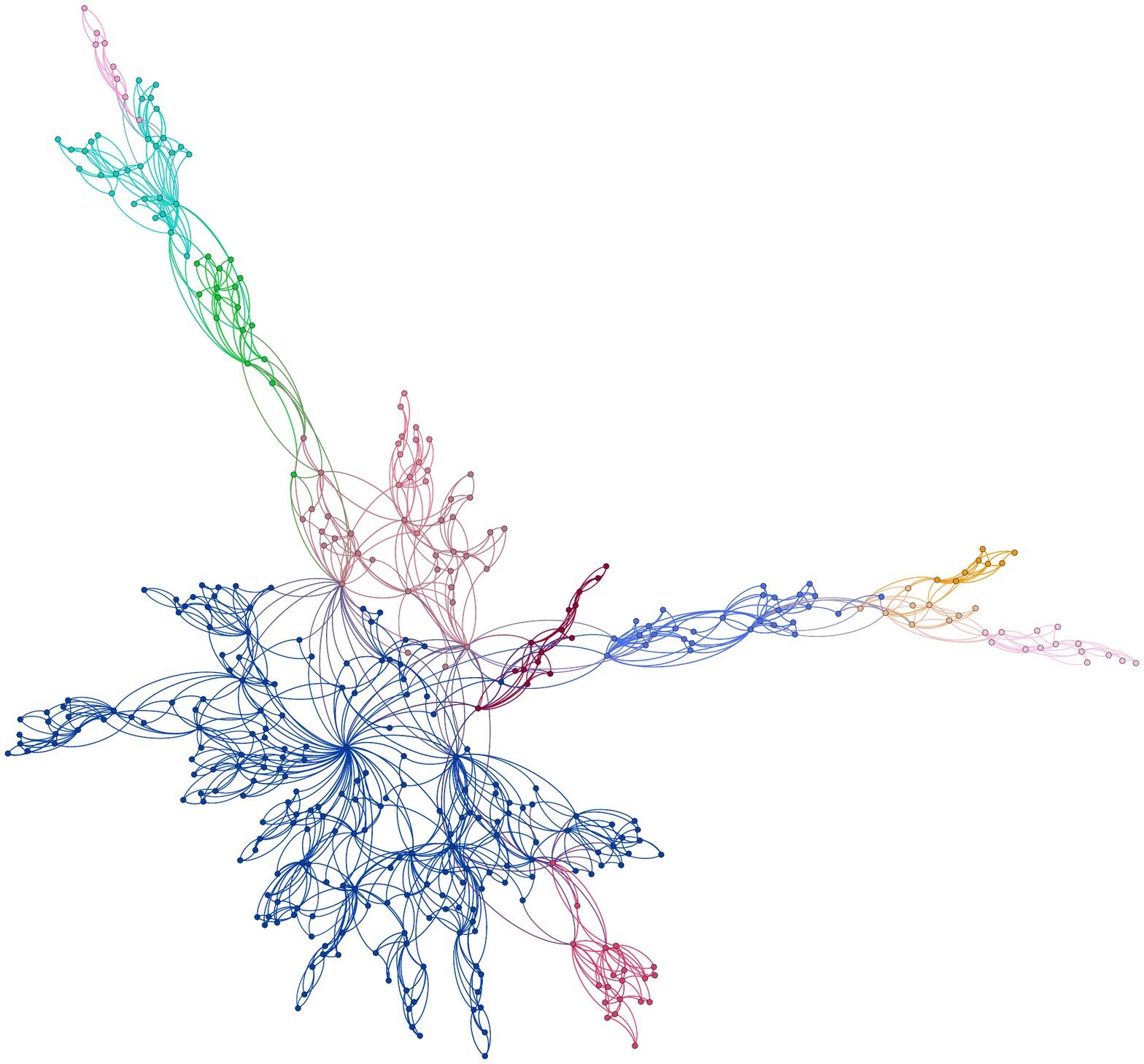}
      \put (33,25) { \textbf{\contour{white}{1}}}
        \put (38,42) { \textbf{\contour{white}{2}}}
         \put (45,34) { \textbf{\contour{white}{3}}}
         \put (57,36) { \textbf{\contour{white}{4}}}
       \put (35,50) { \textbf{\contour{white}{5}}}
             \put (52,7) { \textbf{\contour{white}{6}}}
             \put (20,60) { \textbf{\contour{white}{7}}}
                 \put (76,38) { \textbf{\contour{white}{8}}}
                 \put (82,42) { \textbf{\contour{white}{9}}}
                   \put (13,74) { \textbf{\contour{white}{10}}}
                   \put (9,83) { \textbf{\contour{white}{12}}}
                   \put (86,35) { \textbf{\contour{white}{11}}}
  \end{overpic}
         \includegraphics[width=0.85\linewidth, trim= 0 280 0 250, clip=true]{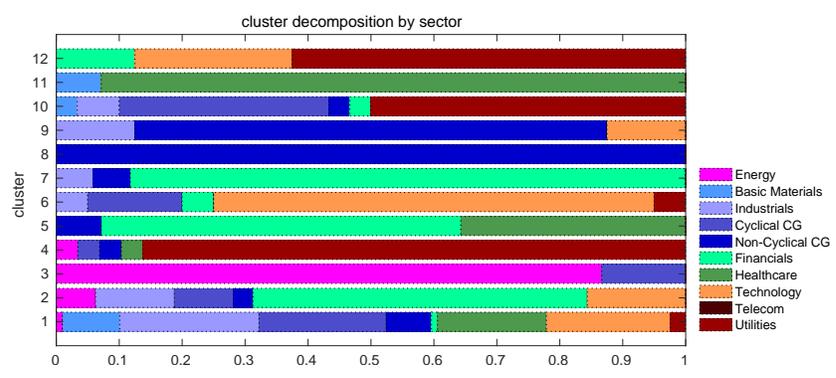}
        \end{center}
        \caption{Clusters in the PMFG.}{\small The network on top shows the color-coded clusters in the PMFG that are idetified by the DBHT algorithm. The bar plot in the bottom shows the decomposition of each of these 12 clusters into fractions, i.e., the share of stocks from a particular sector.}\label{fig:dbht}
\end{figure}

When we look at the MST- and PMFG-filtered networks in figure \ref{fig:pmfg} we note that some clustering emerges naturally. In the following we therefore present a clustering that is based on the PMFG, the so-called DBHT method (short for Directed Bubble Hierarchical Tree).
This particular algorithm produces a clustering that is deterministic 
and does not use any thresholding values and/or parameters and it does not fix the number of clusters a priori.

In short, the  DBHT method \citep{dbht} can achieve a clustering of the PMFG, because the network can be hierarchically divided by separating 3-cliques of nodes. This results in a set of planar graphs, which are referred to as bubbles. Nodes typically belong to more than one bubble, hence in a first step nodes are assigned to a unique bubble. In a second step the algorithm will then merge some bubbles to larger structures based on different distance measures. For further details on the method see also \cite{dbht1} and \cite{dbht2}.

 The results are shown as a color-coded graph in figure \ref{fig:dbht}. We  obtain 12 clusters with varying size. The borders of the groups follow the hierarchical structure of the PMFG.

The grouping is in some relationship with the sector classification, as the bar plot in the bottom of the figure shows. Many clusters are dominated by stocks from only one or two sectors. An exception is the largest cluster (1) which is comprised of a large number of stocks from different sectors which do not show very strong sector-specific behaviors.

\section{GARCH models}\label{sec:garch}

\subsection{Overview}

The stylized facts of financial data presented in section \ref{sec:data} have of course not only inspired research from statistical physicists, but are in fact at the root of the development of financial econometrics as we know it today. Modeling the volatility of an asset serves as the foundation for applications not only in asset pricing but also in risk management, derivatives pricing and forecasting.

The most popular class of models in this field are the GARCH (Generalized Autoregressive Conditional Heteroskedasticity) models. These models are first of all time-series models. They provide a framework to model and estimate the heteroscedasticity in volatility and at the same time they  allow a rudimentary forecast of volatility. They have however originally not been invented with the analysis of correlations in mind, at least not of large samples. Hence, while an estimated dynamic conditional covariance matrix is the natural by-product of any multivariate GARCH model, only recent improvements on the estimation methodology have enabled us to utilize these kind of models in network settings. Independent of this fact, GARCH models, even in simpler form, can be helpful for constructing financial networks, since they allow us to disentangle joint volatility shocks from idiosyncratic ones.

We will first have a look at the univariate GARCH model and some examples of estimated stock volatility. We will then briefly show the basics of a multivariate GARCH and its complications before finally sketching out a network application of such a multivariate GARCH model. 

\subsection{Univariate GARCH}

Let us first have a look a the univariate form of GARCH \citep{Boll}. We assumes that the returns follow a random process with \mbox{$\varepsilon_t = v_t \sqrt{h_t}$},
where $v_t$ is white noise and
\begin{equation}\label{eq:g11}
 h_t = \alpha_0 + \sum_{i=1}^q \alpha_i \varepsilon_{t-i}^2 + \sum_{i=1}^p \beta_i h_{t-i} \; .
\end{equation}

Hence, we assume that the volatility $h_t$ depends on the lagged values of the returns and its own lagged values ($\alpha_i$ and $\beta_i$ are the estimated parameters for lag $i$, $\alpha_0$ is a constant).

\begin{figure}[htb]  
        \begin{center}
      \begin{overpic}[width=\linewidth, trim= 10 270 20 260, clip=true]{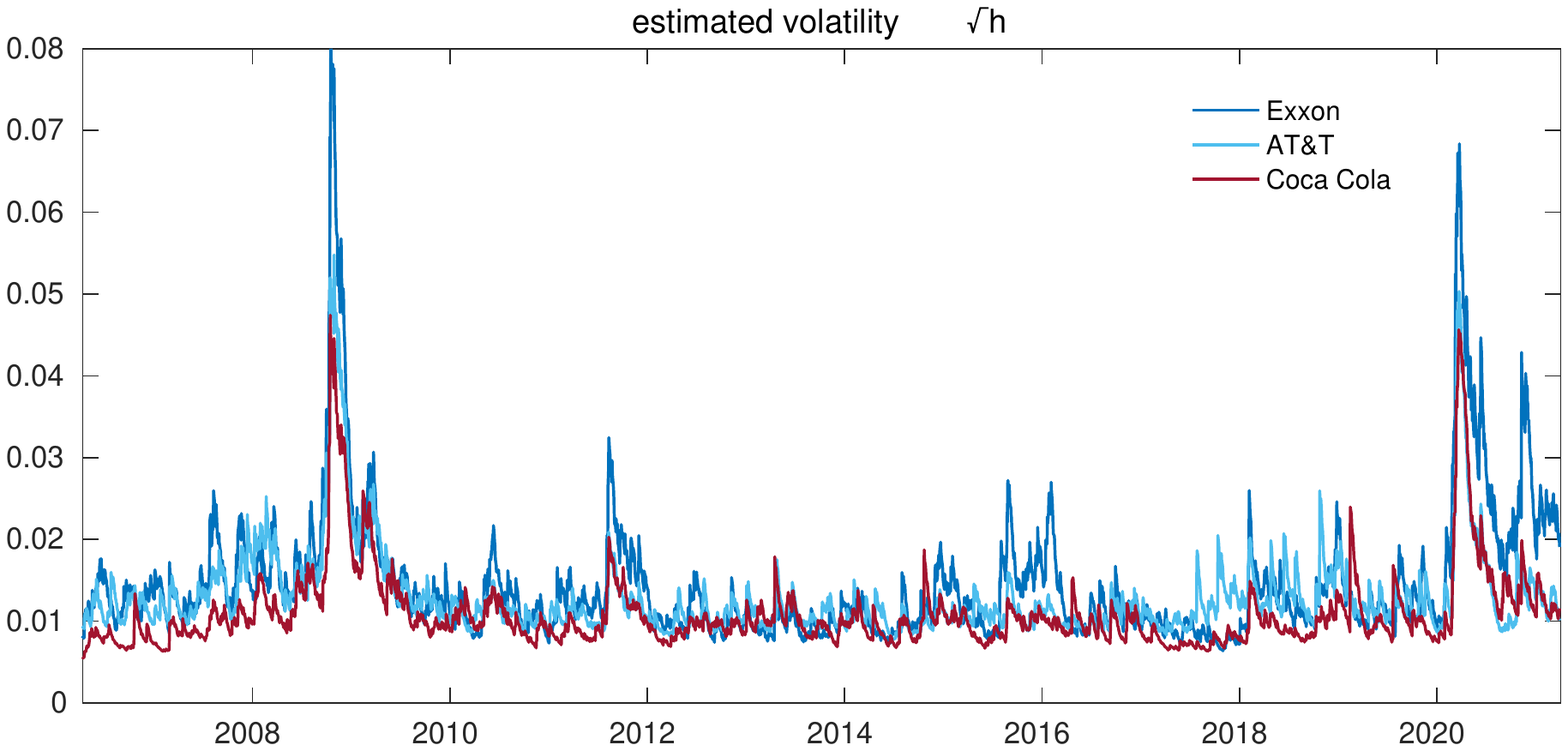}
      \put (50,-2){\scriptsize time}
      \put (2,21){\rotatebox{90}{\scriptsize $\sqrt{h}$}}
      \end{overpic}
        \end{center}
        \caption{Estimated volatility from a GARCH(1,1) model for three stocks.}{\small These three time series of volatility show typical qualitative features, including pronounced peaks and decay. Nevertheless there are significant differences between the stocks that are relevant for the evaluation of stock comovement.}\label{fig:ugarch}
\end{figure}

It is common to refer to a specific lag-specification as a GARCH(p,q)-process. One typically finds that already one lag for $q$ and $p$ are sufficient to describe the volatility by the time series $h_t$ \citep[see also][]{hanslund}. The parameters for this recursion formula can easily be estimated by maximum likelihood. One of the applications of this model is obviously to relate the current (expected) volatility to past realizations of the returns. Another one is that $h_t$ can  be used to normalize (de-garch) the returns and to derive time series with constant unit volatility. Some examples for volatility estimates are shown in figure \ref{fig:ugarch}. Note that most of the GARCH literature differentiates between the actual returns $r_t$ and the process that is modeled, $\varepsilon_t$, like in the equation above. In the following we will however label any return again \mbox{as $r_t$}.

\subsection{Multivariate GARCH}

The GARCH approach has of course been extended to a large number of multivariate versions, the most popular ones are the BEKK \citep{en_kron} and DCC \citep{DCC} model. The volatility is in this case described by a matrix of conditional covariances $H_t$ (and its mean $\bar{H}$). The recursion of the $K$-component BEKK(1,1,K) model can for example be stated as follows:

 \begin{equation}\label{eq:rec}
 H_t=C' C + \sum_{k=1}^K A'_k  \bigl( r_{t-1} r_{t-1}' \bigr)  A_k +
         \sum_{k=1}^K B'_k  H_{t-1}  B_k \;.
\end{equation}

$C, A_k$ and $B_k$ are $N \times N$ matrices ($C$ is upper triangular, we denote the transpose of matrices and vectors by the superscript $ ' \;$).
This specific form has the advantage that it ensures positivity of $H_t$ even for $K=1$. We will not go into the details of multivariate GARCH here, instead the reader is referred to the excellent review by \cite{bauwens}, yet we will discuss some of the problems related to large sample sizes \citep[see also][]{rwgarch} and possible applications to financial networks.

\begin{figure}  
\begin{center}
\begin{overpic}[width=\linewidth, trim= 40 160 40 100, clip=true]{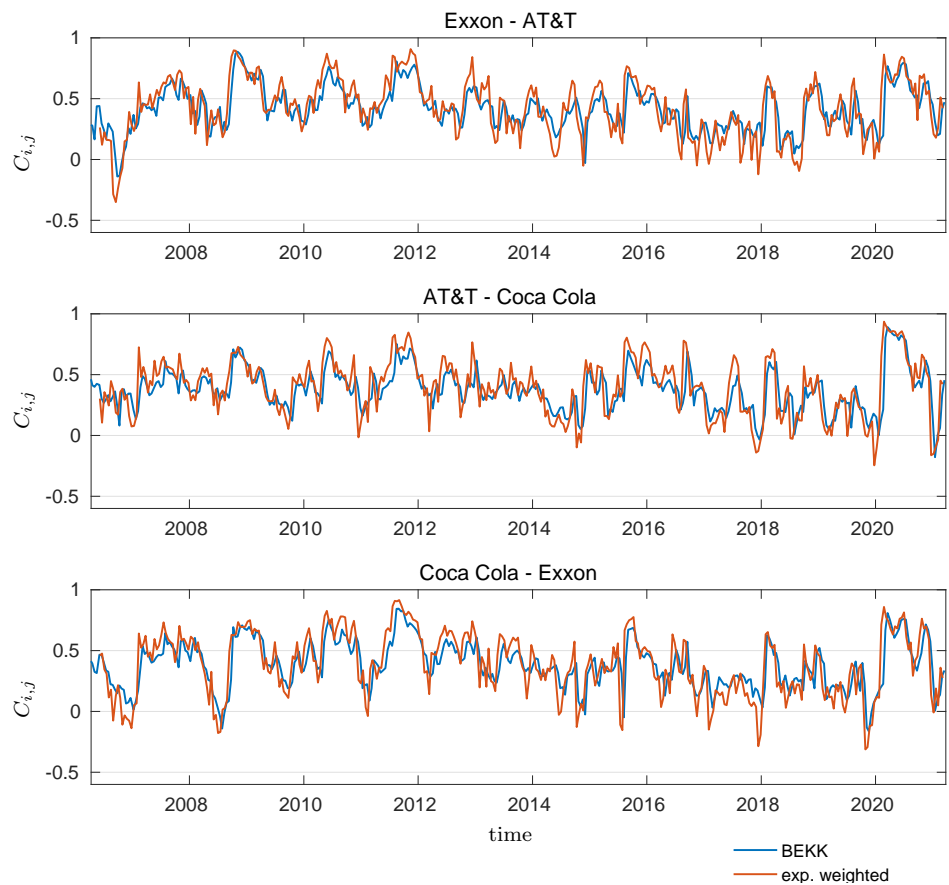}
         \put (3,69){\rotatebox{90}{\scriptsize $C_{i,j}$}}
         \put (3,44){\rotatebox{90}{\scriptsize $C_{i,j}$}}
         \put (3,16){\rotatebox{90}{\scriptsize $C_{i,j}$}}
     \put (49,4){\scriptsize time}
     \end{overpic}
        \end{center}
        \caption{Stock correlations over time.}{\small The plots show comparisons of the correlations estimated with a multivariate GARCH model (using a 10-day average) and those calculated from using the method of exponentially weighted correlations (with a time window of \mbox{$\Delta t=60$} days and $\theta=20$).}
\end{figure}

To start with, when the number of stocks $N$ is large, the high number of so-called nuisance parameters in $\bar{H}$ can lead to biases in the maximum likelihood 
estimation. More important in our case however is the estimation of the dynamics of $H_t$, which runs into similar problems for large samples. Most models therefore try to limit the number of parameters that describe the covariances of assets. For example, in the VEC model
\citep{vbeta} the elements of $H_t$ are treated as a $N(N+1)/2$ vector. In its scalar version \citep{svec} the dynamics are reduced to two 
parameters. The $K$-factor BEKK model \citep{en_kron} shown above uses $N$-dimensional matrix multiplication. 

Additionally, a computational problem is that most estimation methods require $H_t$ to be inverted repeatedly during the estimation. This can further limit the application to large samples. This problem can  partly be circumvented by the use of covariance targeting \citep{EngleMez}.  Here $\bar{H}$ is replaced by the observed time-averaged covariance. This method also allows for covariances and GARCH parameters to be estimated in two separate steps.

Several approaches have worked on the scalability of the estimation of multivariate GARCH models, most notably \cite{aielli} who proposed improvements in the modeling of the conditional covariance matrix and \cite{pakel} who developed an approach that uses a composite likelihood and builds on the estimation of pairs of assets.
A different solution is the method proposed by \cite{large_dcc2} which renders improved estimates by using nonlinear shrinkage.

A different approach to deal with large samples and to reduce the number of parameters 
is to use factor models. An early example is the OGARCH model by \cite{Alex} which allows to drastically reduce the number of parameters by describing the cross-section of asset returns by a number of factors $K$ that can typically be chosen much smaller than $N$. This reduces the number of parameters that have to be estimated substantially.
A more involved variant is the GO-GARCH model \citep{GOGarch}.
 It is applied to the de-correlated returns. The 
principal components of these are then rotated by a $K-$dimensional rotation.  The
$K(K-1)/2$ angles are additional parameters in the estimation.

Yet another approach to reduce the number of parameters is to restrict the form for $H_t$ as in the DECO model \citep{deco}. It assumes identical yet time-varying correlations between all pairs of stocks and one calculates the covariance matrix based on univariate de-garched returns. An alternative way to restrict $H_t$ is presented by \cite{rwgarch}, they employs a dynamic two-factor model and a system of recursion equations to reduce the dynamics.

\subsection{Application}
 
An interesting approach to apply the GARCH framework in a network setting has been presented by \cite{isogai}. He derived dynamic networks of 50 stocks in the Japanese market using the DCC framework. As mentioned above, this framework is advantageous for large $N$ since it can be estimated in two steps. A drawback is that by construction the conditional correlations all follow the same dynamics.

Hence, in this case we model the vector of returns $r_t$ as

\begin{equation}
r_t = \mu_t + H^{1/2}_t z_t \; ,
\end{equation}

where the residuals $\varepsilon_t = H^{1/2}_t z_t$,  $\; E(z_t)=0$ and $Var(z_t) = I_N$. The $N \times N$ matrix $H_t$ corresponds to
 the conditional covariance of $r_{t}$. $H_t^{1/2}$ is positive definite.
 
 The mean $\mu_t$ is modeled as an autoregressive moving average model  depending on lagged returns and volatility. The volatility part is given by
 
  \begin{equation}
 \tilde{H}_t = \Omega + \sum_{i=1}^q S_i \varepsilon_{t-i} \odot \varepsilon_{t-i}  + \sum_{j=1}^p T_j \tilde{H}_{t-j} \; , 
\end{equation}

where  $\tilde{H}_t$ is the diagonalized version of $H_t$.  $S$ and $T$ are also diagonal matrices, $\odot$ is the Hadamard (element-wise) product. The actual dynamic correlation  $R_t$ of the model is calculated via a proxy variable $Q_t$ (that again depends on the lagged $z$ and $Q$) to ensure that $R_t$ stays positive definite, which leads to the well known sandwich structure where

  \begin{equation}
 R_t = diag(Q_t)^{-1/2} \; Q_t \; diag(Q_t)^{-1/2} \; .
\end{equation}

These dynamic correlations can then be used to describe the relationships between stocks over time. \cite{isogai} presents that these correlations will in many cases differ from correlations calculated in rolling windows applied to $r_t$.

\section{Pairwise estimation of dependencies}\label{sec:pair}

\subsection{Estimating feedback}

A seminal work in the literature on the estimation of dependencies of time-series is the paper by \cite{granger69}. This paper discusses, in a rather general sense, approaches to identify influences of co-evolving processes, also called feedback. The idea that became famous however was the two-variable model that describes the influence of two stationary time series $x_t$ and $y_t$ (stationary and with zero mean), which can be modeled by their lagged values, hence

\begin{align}
x_t = \sum_{j=1}^m a_j x_{t-j} +  \sum_{j=1}^m b_j y_{t-j} + \varepsilon_t \; \; \\
y_t = \sum_{j=1}^m c_j x_{t-j} +  \sum_{j=1}^m d_j y_{t-j} + \eta_t \; ,
\end{align}

where $ \varepsilon$ and $\eta$ are uncorrelated noise.

In this case the influences from $x$ to $y$ and vica versa are described by the parameters $a$, $b$, $c$ and $d$. Today we would refer to this specification as a vector autoregressive model (an alternative would be to allow for contemporaneous effects). When we restrict the model to the case $j=1$ we arrive as what is often stated as a model that test \emph{Granger causality}. This expression should not be taken too literally. It refers to the idea that if the values of one time-series explain the lagged values of the other, then this temporal relationship might also indicate some fundamental relationship between the variables \citep[see also][]{zellner62,luet91}.

\subsection{Application of Granger-causality}

There have been several applications of Granger-causality to financial time series, yet the most interesting and complete one in the context of financial networks is probably the one by \cite{Billio2012}. They analyze Granger-causality for a  sample of 100 hedge funds, brokers, banks and insurers for the the time before and during the 2008 financial crisis.
For the time series of returns $i$ and $j$ they test the significance of their connection by estimating $a_i$, $a_j$, $b_{ij}$ and $b_{ji}$, hence

\begin{align}\label{eq:billio}
r_{i,t} &= a_i r_{i,t-1} +  b_{ij}  r_{j,t-1} + \varepsilon_{i,t} \;   \\
r_{j,t} &= a_j r_{j,t-1} +  b_{ji}  r_{i,t-1} + \varepsilon_{j,t} \; .
\end{align}

If the $b$-parameters are significantly different from zero, one can speak of a relationship between $i$ and $j$. This relationship can be uni-directional or bi-directional.

There are two aspects to note for this model. First, it is not at all likely to have significant relationships that explain the return of another asset in the future (the study uses monthly returns). One would expect that arbitrage and a certain level of market efficiency would prohibit this. In fact the study by \cite{Billio2012} shows that at most times the number of connections that is significant at a 5\% level is only slightly larger than 5\%. This however changes in times of pronounced market stress, in this case at the end of 2007. Hence, while this method could show the effects of the financial crisis relatively well, applications to other time periods can be more challenging.

The second aspect is of course related to the long memory in asset prices. Similarities in volatility changes of the assets would drastically influence the estimation results for all parameters and would lead to a misrepresentation of connectivity. The analysis is therefore not performed with the raw returns but the authors assume a GARCH(1,1) model. In practice we can assume that the returns have been de-garched (see equation \ref{eq:g11}) before estimating the parameters for the model in equation \ref{eq:billio}. 

\subsection{Pairwise estimation with a larger data set}

A related approach to measure financial dependencies has been proposed by \cite{interc}. Here the estimation is based on synchronous daily and weekly returns of a much larger sample of stocks from different markets. Also here we assume that the returns follow a random process with \mbox{$\varepsilon_t = v_t \sqrt{h_t}$},
where $v_t$ is white noise and $h$ follows a GARCH(1,1) process.

The conditional variance $h_t$ is then used to produce filtered returns $r^f$, such that 
\begin{equation}
r_{i,t}^{f} = \frac{r_{i,t}} {\sqrt{h_{i,t}} } \; .
\end{equation}

By de-garching the returns we obtained time series that can be treated in an almost standard regression framework, since performing  pairwise regressions of all the filtered returns generates a measure for the comovement. It should however be noted that the distributions of the filtered returns are not perfectly normally distributed. It is therefore appropriate to assume t-distributed errors and apply what is mostly called the robust regression approach \citep{robust} (the distributions of filtered returns are closer to a Normal distribution for monthly returns).

Hence, to measure dependence we estimate the dependencies for all pairs of stocks $i,j$ as 

	\begin{equation}
	r_{i,t}^{f} = \beta_{0,ij} + \beta_{1,ij} r_{j,t}^{f} + \varepsilon_t \; .
	\end{equation}
	
The results of the estimation can now be mapped into a network representation.	
We choose a threshold in terms of the significance level and construct networks in which connections are defined by observing a significantly positive $\beta_{1,ij}$. For constructing weights we can use the corresponding p-values, namely $p_{ij}$ (of the null-hypothesis significance test for $\beta_{1,ij}$), that are obtained in the course of the estimation. The exact values of the link weights $A_{ij}$ correspond to the extend to which $p_{ij}$ exceeds a given threshold $\gamma$, i.e.

\begin{equation}
A_{ij} \propto (\gamma - p_{ij}) \; .
\end{equation}

This adjacency matrix has a weighted positive entry if stocks $i$ and $j$ are significantly linked, measured by the estimated conditional correlation.

\begin{figure}[p] 
        \begin{center}
        \includegraphics[width=0.94\linewidth, trim= 0 53 10 57, clip=true]{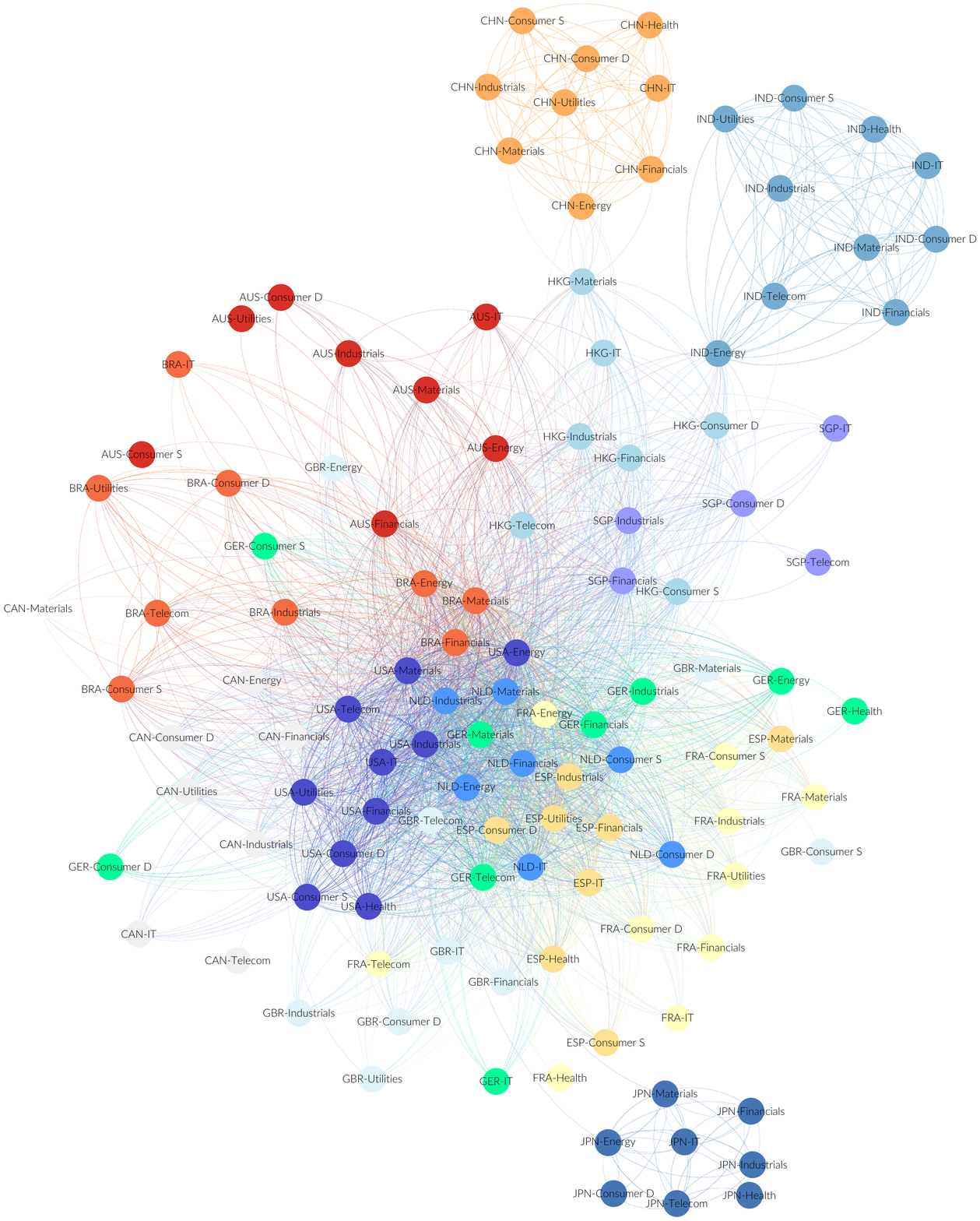}
        \end{center}
        \caption{Vizualization of pairwise estimated stock dependencies, averaged on the sector level, color-coding applied by country.}{\footnotesize Stocks from the materials, energy and financial sector, mainly from the US (purple), Germany (green), the UK (gray), The Netherlands (blue), France (yellow) and Spain (beige), form the densly connected center of this network. The markets of Canada, Brasil, Australia, China, India, Singapore and Hong Kong show in mostly peripheral positions.}\label{fig:conn}
\end{figure}

In this particular case where the assets under consideration come from different markets from all around the globe  the time difference in trading hours are an additional complication. The estimated  dependencies based on daily returns  may be  biased downwards because the time series are asynchronous \cite[see also][]{martens,christensen,hayashi}.
 In general, this can be dealt with in two ways: by using tick data and calculation of synchronous pseudo closing prices, or by time aggregation. The approach by \cite{interc} uses an approximate correction for this effect that is applied to the p-values derived from the estimation.
 The method assume that the estimation results based on the weekly data are unbiased. The results from the daily data are thus adjusted in such a way that they are consistent with the dependencies obtained from the weekly data.
 
Since the sample is rather large it is for most purposes not useful to report results only on the level of stocks. A dimensional reduction can be achieved by hierarchically aggregating the results on the level of sectors and countries (according to GICS and TRBC classification).
The resulting network is shown in figure \ref{fig:conn}. The p-values have been averaged on a sector-by-sector and country-by-country level.
 The visualization shows that many sectors of `western' markets are globally integrated, while some markets like China, Japan and India remain more segregated.
  
\section{Variance decomposition}\label{sec:var}

In the previous section we have already touched upon the use of vector autoregressive models (VAR) for the analysis of dependencies in financial markets. While many applications of such models to financial markets and macroeconomic performance exist, these have mostly been cases with samples of limited size.
One reason is that once $N$ becomes large it can become difficult to estimate all possible influences on a certain time-series jointly.  
The other reason is that one is typically interested in obtaining an orthogonalized version of the shocks to calculate impulse responses or decomposed forecast errors, which can become troublesome even for moderate $N$ \citep[see also][]{koop96,pes1,pes2}.

A decisive improvement on this issue is the variance decomposition proposed in \cite{yilmazvar}.
As in any case, we will first estimate a standard VAR model for a specific sample or time window with a certain number of lags. Then we can evaluate in how far each variable contributes to the generalized forecast error variance decomposition of the other variables. For the application shown in \cite{diebold2014network}, the idea is that the sums of these contributions can then for example be used to calculate  contributions to market volatility and are also seen as a proxy for contributions to systemic risk.

We assume that the underlying process $x_t$ that describes, for example, a vector with daily volatility of stocks, follows $x_t = \Theta (L) u_t$, where 

\begin{equation}
\Theta (L) = \Theta_0 + \Theta_1 L + \Theta_L^2 \; .
\end{equation}

Hence, $x_t$ depends on the lagged orthogonal shocks $u_t$ in a way that is described by the parameters contained in the $\Theta$ matrices. $L$ is the lag operator. The decisive part of this approach is now to calculate the generalized variance decomposition matrix $D^H$ with entries $d^H_{ij}$ that specify how stock $i$'s $H$ step ahead forecast error variance is influenced by stock $j$ (where $i\neq j$). These entries are given by

\begin{equation}
d^H_{ij} = \frac{\sigma^{-1}_{jj}\sum_{h=0}^{H-1}(e'_i \Theta_h \Omega e_j)^2}
 {\sum_{h=0}^{H-1} (e'_i \Theta_h \Omega \Theta'_h e_j)} \; ,
\end{equation} 

where $e_j$ is a vector where the $j$th element is 1 and the others are zeros. $\Theta_h$ is the coefficient matrix multiplying the $h$-lagged
shock vector in the infinite moving-average representation of
the non-orthogonalized VAR. $\Omega$ is the covariance matrix of the shocks and $\sigma_{jj}$ is the $j$th element of the diagonal of $\Omega$.

This variance decomposition matrix can easily be interpreted as an adjacency matrix. The matrix is directed, because the elements $d_{ij}$ and $d_{ji}$ are typically not the same. The sums over the rows of this adjacency matrix sum to 1 since they resemble the shares of received contributions to the variance decomposition. To account for the remainder of these contributions it makes sense to write the diagonal elements as $A_{ii} = \sum_{j,j\neq i} A_{ij}$.

While this approach presents a useful connection from the class of VAR models towards network models, there are also some weaknesses. As \cite{chanlau17} and \cite{lanne} point out, the generalized forecast
error variance decomposition in the form presented in the original paper can lead to inconsistencies when comparing the contributions of firms to systemic risk over time. Improvements in the method can however mostly mitigate this issue. Another issue is the still somewhat limited scaleability with respect to the sample size, which is discussed by \cite{largevar}.

A further development in this class of models is the so-called TVP-VAR (time-varying parameter vector autoregressive model).
Here the variance-covariance matrix is allowed to vary by applying a Kalman
filter estimation with forgetting factors. This means that it is no longer necessary to choose time windows and window sizes. For the details the reader may refer to \cite{forget_koop,luetpet,koryil,anton1} and \cite{anton2}.

\section{Conclusions}\label{sec:conc}

This collection has shown that a variety of viewpoints exist in analyzing financial dependencies. While some approaches focus on topological network representation constraints, others are more focused on a modeling of details of the time-series properties of returns. This has of course implications for the scalability to larger data sets and its applicability to different current issues in financial market research. Nevertheless, this survey has also shown that the field has moved beyond a phase of exploration of asset market data. 

Therefore we encourage researchers to take into account results outside of their domains and to apply network-related research to problems that are relevant from a societal, economic or political perspective. We believe that a collaborative effort among different disciplines will be the key to solve many current challenges.

%

\bibliographystyle{elsarticle-harv} 

\bibliography{network2}

\appendix
\newpage
\section{Correlation-based graphs}
\begin{figure}[hbt!]
\begin{center}
        \includegraphics[width=0.45\linewidth, trim= 10 100 10 100, clip=true]{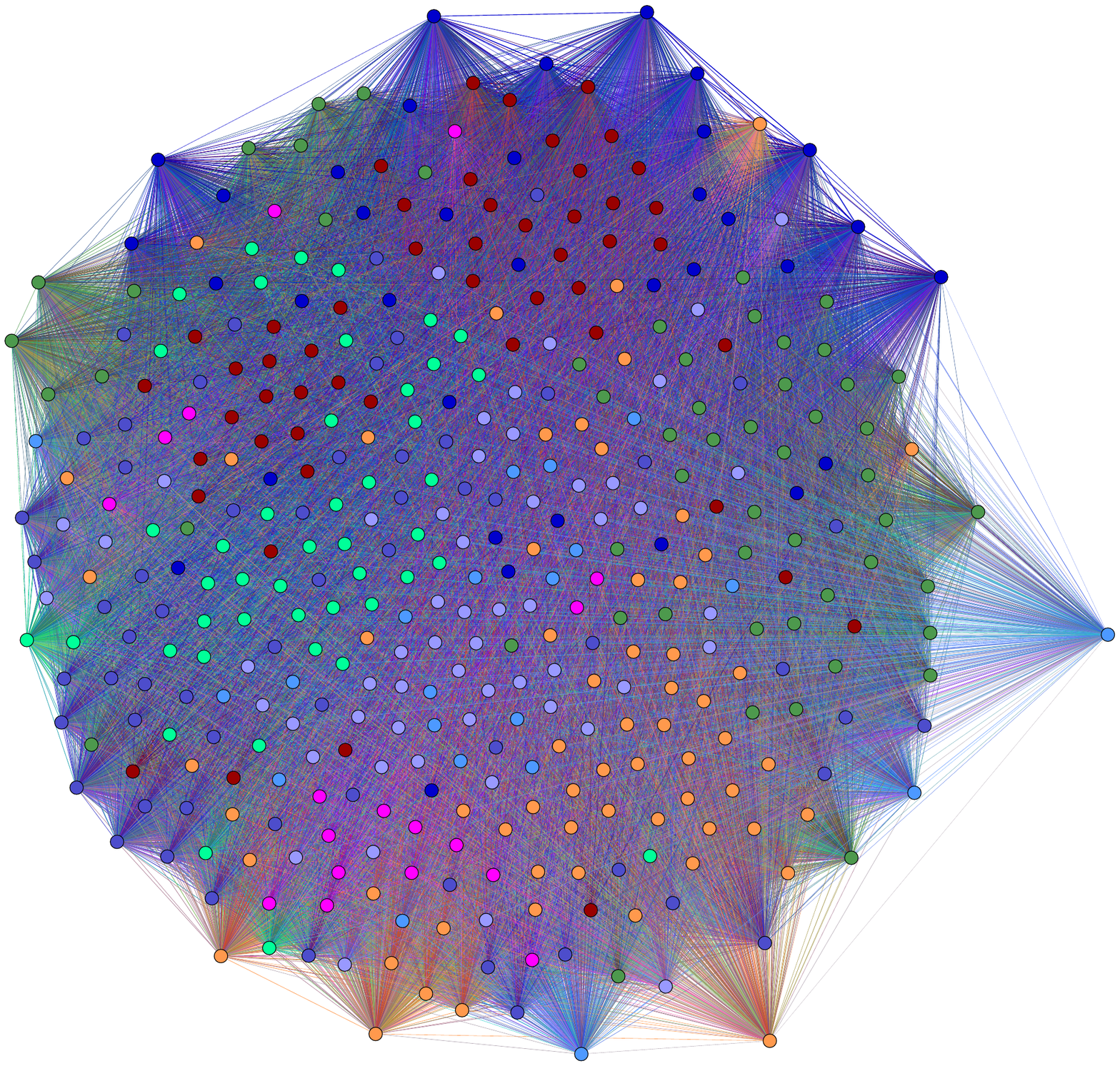}
         \includegraphics[width=0.45\linewidth, trim= 10 100 10 100, clip=true]{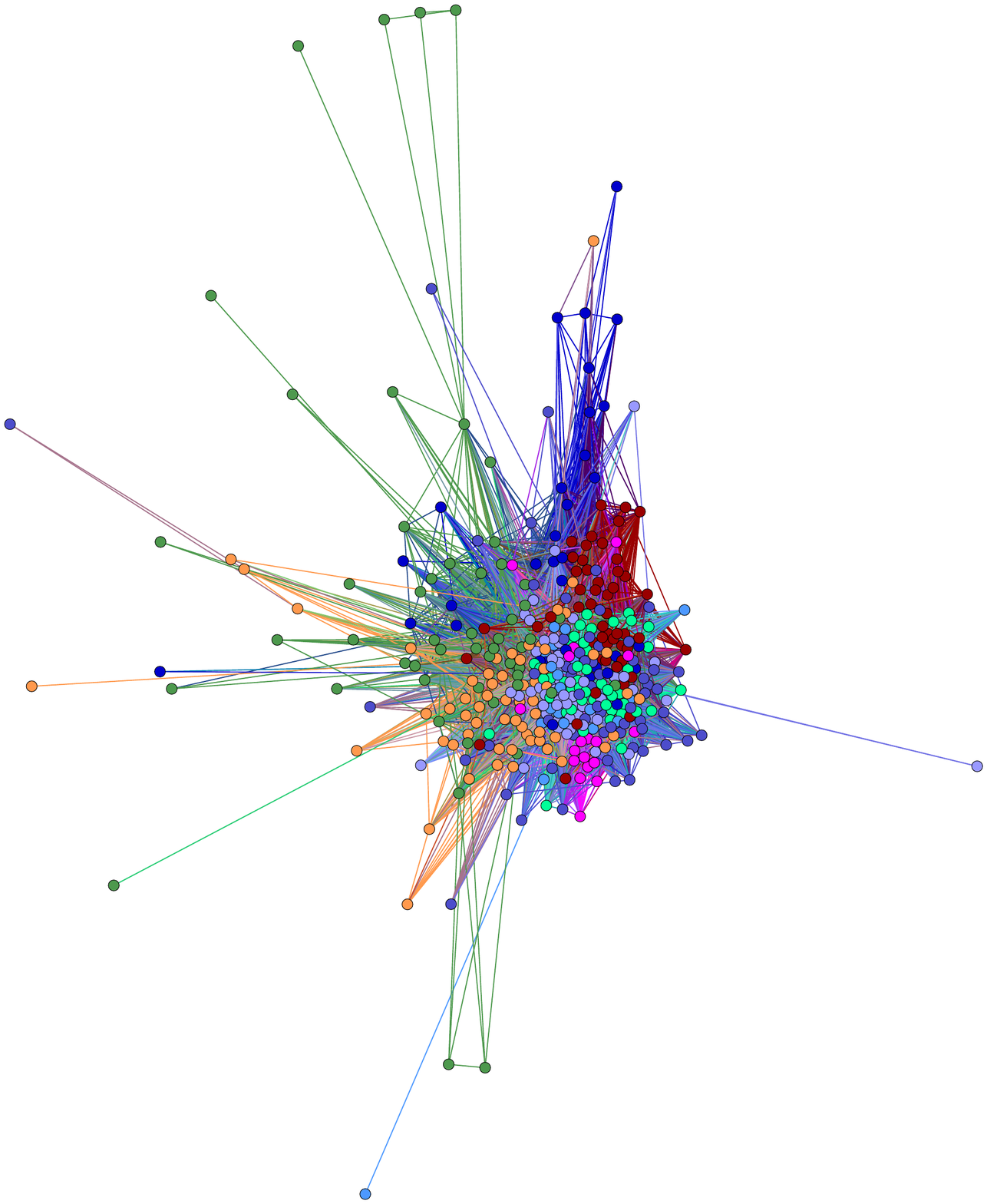}
          \includegraphics[width=0.45\linewidth, trim= 10 110 10 110, clip=true]{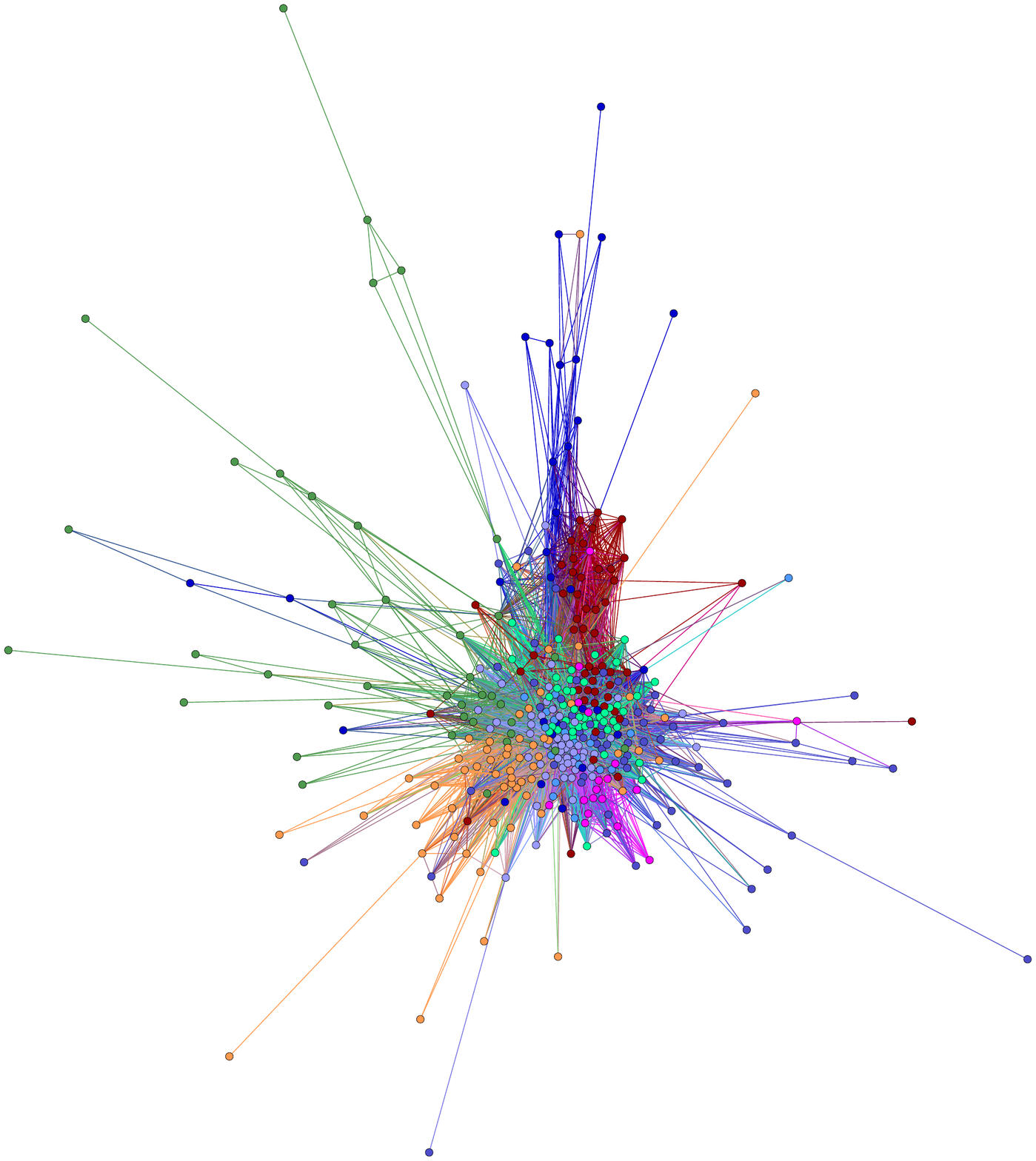}
           \includegraphics[width=0.45\linewidth, trim= 10 110 10 110, clip=true]{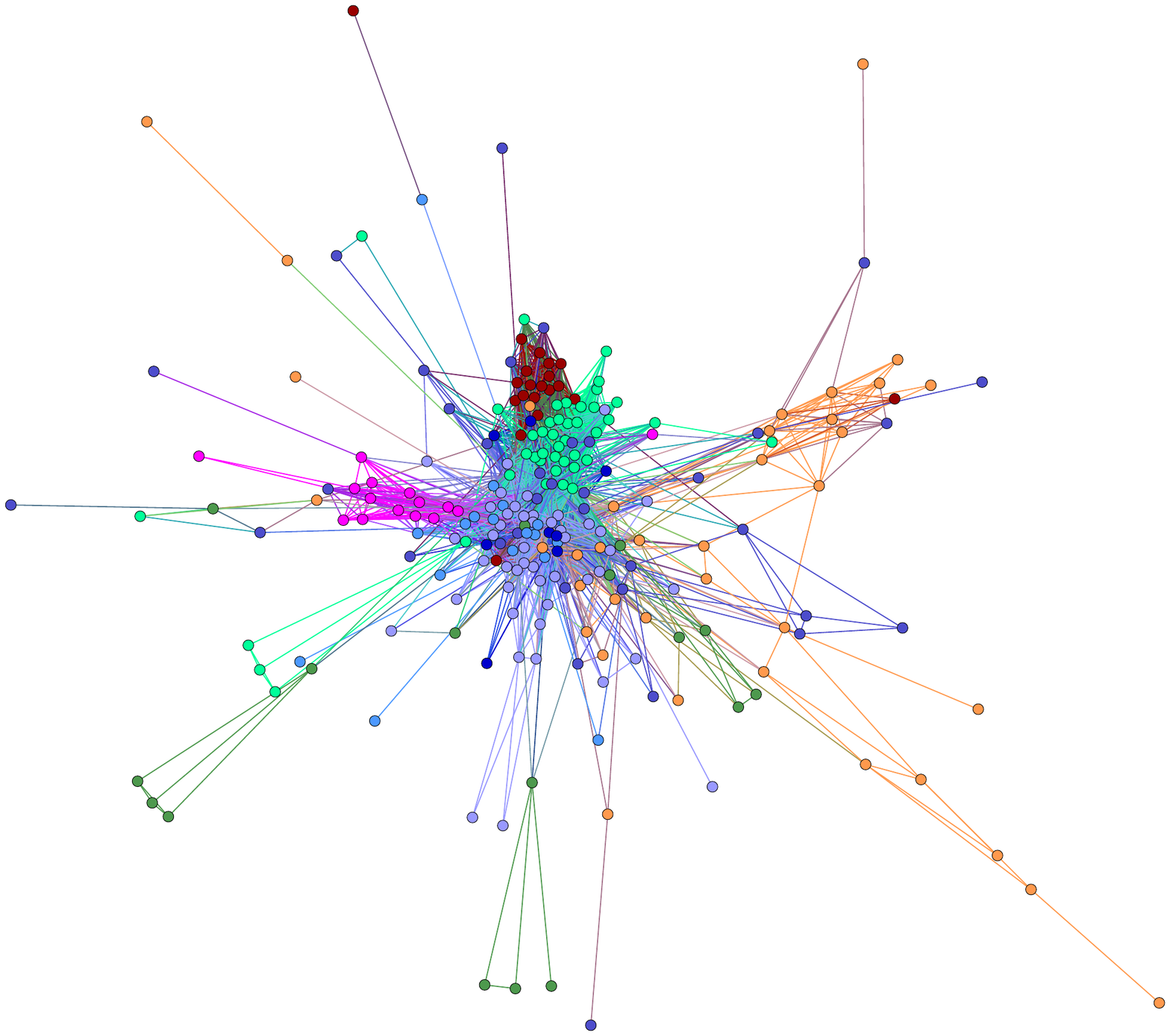}
\end{center}
        \caption{Complete network and networks with links above the 60\%, 80\% and 95\% interval of the distribution of correlation coefficients.}{\footnotesize The giant connected components consist of 404, 397, 372, 240 nodes and 81,406, 32,562, 16,279, 3,727 links. Color-coding by sector classification: pink - Energy, bright blue - Basic Materials, bright purple - Industrials, dark purple - Cyclical Consumer Goods, dark blue - Non-cyclical Consumer Goods, bright green - Financials, dark green - Healthcare, orange - Technology, brown - Utilities. }\label{fig:corrbased2}
        
\end{figure}

\end{document}